\documentclass[twocolumn,pra,showpacs,amsmath
 ,superscriptaddress]{revtex4-1}
\usepackage{epsfig,graphicx}

\newcommand{\be}{\begin{equation}}
\newcommand{\e}[1]{\label{#1}\end{equation}}
\def\bea{\begin{eqnarray}}
\def\ea#1{\label{#1}\end{eqnarray}}
\def\rqn#1{(\ref{#1})}
\def\ee{\end{equation}}
\def\eea{\end{eqnarray}}
\def\bes#1{\begin{subequations}\label{#1}}
\def\ese{\end{subequations}}

\begin{document}

\title{Connection-state approach to pre- and post-selected quantum measurements}
\author{Abraham G. Kofman}
 \affiliation{CEMS, RIKEN, Saitama, 351-0198, Japan}
\author{\c{S}ahin K. \"{O}zdemir}
 \affiliation{CEMS, RIKEN, Saitama, 351-0198, Japan}
\affiliation{Washington University, Saint Louis, Missouri 63130-4899, USA}
\author{Franco Nori}
 \affiliation{CEMS, RIKEN, Saitama, 351-0198, Japan}
\affiliation{Physics Department, The University of Michigan, Ann Arbor, Michigan 48109-1040, USA}

\date{\today}

\begin{abstract}
We discuss the concept of connection states (or connection matrices) that describe posterior ensembles, post-selected according to the outcomes of a quantum measurement.
Connection matrices allow one to obtain results of any weak and some non-weak pre- and post-selected measurements, in the same manner as density matrices allow one to predict the results of conventional quantum measurements.
Connection matrices are direct extensions of the density matrices and are generally non-Hermitian, which we show to be a direct consequence of quantum complementarity.
This implies that  the ultimate reason for unusual weak values is quantum complementarity.
We show that connection matrices can be determined experimentally.
We also show that retrodictive states are a special case of connection states.
We propose a new method of tomography of quantum detectors.
\end{abstract}

\pacs{03.65.Ta, 03.65.Ca, 03.65.Wj}
\maketitle

\section{Introduction}

Quantum mechanics is intrinsically stochastic.
Measurements of a physical quantity for an ensemble of identical quantum systems prepared in the same state (the so-called ``preselected'' ensemble) yield different outcomes for different individual systems.
The quantum state, described generally by the density matrix $\rho$, contains all the information on a quantum system available for an observer at a given time $t_0$ and thus allows one to predict the probabilities of the outcomes for any future measurement performed on the system.

A measurement performed at a time $t_1\ (t_1>t_0)$ allows one to divide the preselected ensemble into pre- and post-selected (PPS) ensembles, i.e., sub-ensembles corresponding to different measurement outcomes.
The information provided by a measurement outcome $l$ is completely described by a POVM operator $E_l$.
Thus, the quantum systems in a PPS ensemble, corresponding to an outcome $l$, are characterized by two quantities, the initial state $\rho$ and the observable $E_l$.
PPS ensembles can be probed by measurements performed at intermediate times $t\in(t_0,t_1)$, i.e., by PPS measurements \cite{aha64}.
The results of PPS measurements are obtained, conceptually speaking, by combining the prediction based on the initial state and retrodiction (retroactive prediction) based on the final-measurement outcome, such a procedure being called ``quantum smoothing'' \cite{tsa09}.

We ask the following question:
Can one describe a PPS ensemble similarly to a pre-selected ensemble, i.e., by some operator, which can be used to obtain the full statistics of any PPS measurement?
Below it is shown that indeed, under certain conditions, such an operator exists.
It is called here ``connection state'' (or ``connection matrix''), since it describes the state of quantum systems between the preparation of the initial state and the measurement and thus connects these two stages of evolution.
The connection matrix is useful for quantum smoothing, in contrast to the density matrix used for prediction.
Remarkably, the connection matrix is generally non-Hermitian.
Below this property is shown to be a direct consequence of the non-classical nature of quantum mechanics. 

PPS measurements \cite{aha64}, especially weak PPS measurements and the resulting weak values of observables \cite{aha88}, were found useful in a multitude of interesting applications (for reviews see Refs.~\cite{aha02,aha05,aha10,kof12,shi12,dre}), including quantum paradoxes \cite{aha91,res04,aha02a,*lun09,*yok09,aha05}, foundations of quantum mechanics \cite{gog11}, measuring wavefunctions \cite{lun11,*sal13,koc11,*bli13}, high-precision metrology \cite{hos08,*dix09,*zho}, and plasmonics \cite{gor12}.
These applications involve unusual properties of PPS measurements, such as unusual weak values, which can be complex numbers with unbounded magnitudes and real parts lying outside the range of the eigenvalues of the observable.

The physical meaning of weak values is not yet completely understood and is a subject of controversy \cite{leg89,per89,sve}.
This is an impediment to further progress in the field of PPS measurements.
In view of the importance of unusual weak values, they have attracted significant attention.
In particular, distributions of unusual weak values were studied \cite{ber10,*ber11}, and some necessary conditions for such values were obtained \cite{aha91,aha02,kof12}.
Note, however, that non-classical behavior in weak PPS measurements is not always directly associated with unusual weak values.
In particular, there are quantum paradoxes, such as the three-box problem \cite{aha91,res04} and ``the quantum-Cheshire-cat paradox'' \cite{aha05,aha}, which are obtained with the {\em usual} weak values.
This shows that the weak values of the measured observables are not the most fundamental quantities characterizing weak PPS measurements.

The connection state provides the full statistics of any weak PPS measurements for a given PPS ensemble.
Therefore,  the connection state is the most fundamental quantity characterizing weak PPS measurements.
Connection states are not only more fundamental but also simpler mathematically than weak values, since generally connection states involve two operators and not three as weak values.
Therefore the analysis of connection states is comparatively easy, which allows us to shed new light on PPS measurements.
In particular, below we show that the unusual character of weak values is explained by the non-Hermitianity of connection states and hence is a direct consequence of the non-classical nature of quantum mechanics.

Some special cases of connection states were introduced previously \cite{shi10,*hos11,hof10} as convenient auxiliary tools.
Here we emphasize the fundamental importance of the connection states and discuss in detail their properties and the physical significance.
After the preprint of the first version of this paper was published \cite{kof13}, the paper \cite{hir13} appeared where a ``transient state'' coinciding with our connection state was considered.
Actually, only the Hermitian part of the connection state is important for the problem discussed in Ref.~\cite{hir13}.
Here both the Hermitian and anti-Hermitian parts of connection states are studied on an equal footing, and the importance of both of them is shown.

The outline of the paper is as follows.
Section \ref{II} provides an overview of conventional and PPS quantum measurements.
In Sec.~\ref{III}, the connection states are introduced in the context of weak PPS measurements.
We discuss properties, physical meaning, the relation to weak values, and tomography of connection states.
In Sec.~\ref{IV}, we show that connection states describe not only weak measurements but also a class of arbitrary-strength PPS measurements.
In particular, we consider violations of uncertainty relations in PPS measurements.
We also show that retrodictive states \cite{bar00} are a special case of connection states; as a result, we obtain a symmetric form of connection states.
Next, we propose a new method for tomography of quantum detectors.
In Sec.~\ref{V}, we show that the unusual properties of the class of PPS measurements described by connection states are a direct consequence of the non-classical nature of quantum mechanics.
In Sec.~\ref{VI}, the effects of the unitary dynamics of quantum systems on connection states and PPS measurements are considered.
In Sec.~\ref{VII}, we discuss the relations between connection and posterior states and also between the present approach and the two-state vector formalism \cite{aha64,aha02,aha10}.
Section \ref{VIII} contains conclusions.
Three Appendices provide details of calculations.

\section{Overview of quantum measurements}
\label{II}

\subsection{Conventional quantum measurements}

We begin with a very brief overview of quantum measurements.
First, we consider conventional quantum measurements.
Let $\rho$ be the density matrix describing the state of the quantum system.
The operator of any physical quantity $A$ has the spectral expansion  
\be
A=\sum_ia_i\Pi_i.
 \e{4}
Here $a_i\ (a_i\ne a_j\ \forall i\ne j)$ are the eigenvalues of $A$, and $\Pi_i$ are projection operators that satisfy the equalities 
\be
\Pi_i\Pi_j=\Pi_i\,\delta_{ij},\quad\sum_j\Pi_j=I, 
 \e{23}
where $I$ is the unity operator.
Then, according to the projection postulate \cite{neu55,lud51}, an ideal (strong) measurement of $A$ yields an eigenvalue $a_i$ with probability given by the Born rule,
 \be
P_i={\rm Tr}\,(\rho\,\Pi_i),
 \e{7}
while leaving the system in the posterior state 
 \be
\rho_i'=\Pi_i\rho\,\Pi_i/{\rm Tr}\,(\rho\,\Pi_i).
 \e{35}
Generally, this state depends on the intial state $\rho$.
However, when $\Pi_i$ is a rank-1 projector, $\Pi_i=|\phi_i\rangle\langle\phi_i|$, the posterior state $|\phi_i\rangle$ is pure and independent of the initial state.

Furthermore, the most general measurements are described by the positive-operator valued measure (POVM) $\{E_l\}$ satisfying  
\be
\sum_lE_l=I.
 \e{71}
Now the probability of the $l$th measurement outcome  is given by \cite{nie00}
\be
P_l={\rm Tr}\,(\rho E_l).
 \e{16}

\begin{figure}[htb]
\begin{center}
\includegraphics[width=7cm]{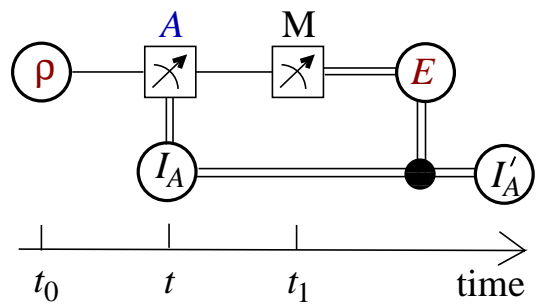}
\end{center}
\caption{Schematic diagram of PPS measurements.
A conventional measurement of an observable $A$ in a state $\rho$ provides information $I_A$.
In a PPS measurement, this information is conditioned on a result of a subsequent measurement M, described by a POVM operator $E$.
This yields modified information $I_A'$ [see, e.g., Eqs.~\rqn{8} and \rqn{2}].
Here ordinary (double) lines carry quantum (classical) information.
}
 \label{f1}\end{figure}

\subsection{Pre- and post-selected measurements}

Consider now PPS measurements (see Fig.\ \ref{f1}).
We begin with ideal (or strong) PPS measurements \cite{aha64}.
In the general case, when the initial state $\rho$ can be a mixed state and the post-selection is made by an outcome of a general measurement with a POVM operator $E$, it is  easy to show with the help of Eqs.~\rqn{7}, \rqn{35}, and \rqn{16} that the probability to observe an eigenvalue $a_i$ of $A$ in a strong PPS measurement is given by \cite{kof12,dre12}
 \be
P_{i|E}\,=\frac{{\rm Tr}\,(E\,\Pi_i\,\rho\,\Pi_i)}{\sum_j{\rm Tr}\,(E\,\Pi_j\,\rho\,\Pi_j)}.
 \e{8}
This formula is an extension of the Aharonov-Bergmann-Lebowitz (ABL) rule \cite{aha64,aha91}, reducing to the latter when the initial state is pure and the post-selection measurement is ideal. 
Studies of strong PPS measurements provided some interesting results, such as the 3-box quantum paradox \cite{aha02,res04} and the time-symmetry relation \cite{aha64,aha02,aha05,kof12}.

An alternative to strong PPS measurements are weak PPS measurements \cite{aha88}.
Weak PPS measurements of a quantity $A$ yield the so-called weak value of $A$, which in the general case is given by \cite{wis02,joh04}
 \be
A_w\:=\:\frac{{\rm Tr}\,(EA\rho)}{{\rm Tr}\,(E\rho)}.
 \e{2}
In the special case when the initial state is pure, $\rho=|\psi\rangle\langle\psi|$, and the final measurement is ideal with $E$ being a rank-1 projector, $E=|\phi\rangle\langle\phi|$, the weak value is \cite{aha88} 
 \be
A_w=\frac{\langle\phi|A|\psi\rangle}{\langle\phi|\psi\rangle}.
 \e{76}
In this case, it is common to say that ``the system is post-selected in the state $|\phi\rangle$'', since the final state is $|\phi\rangle$.
Note, however, that generally the final state is not uniquely determined by $E$ and is dependent on $\rho$ as well.
What is more important, {\em the final state is irrelevant for PPS measurements}, since they are not affected by the evolution of the system after the post-selection.
Still, we will use the common term ``a pure post-selected state'' to refer to the cases where $E$ is a rank-1 projector.

In PPS measurements, the intermediate measurement of $A$ generally significantly modifies the initial state $\rho$, which may result in a change of the post-selection probability, thus affecting the PPS ensemble \cite{bub86}.
In particular, for strong PPS measurements this effect is seen from the fact that the post-selection probability, given by the denominator in Eq.~\rqn{8}, depends explicitly on $A$.
Below we focus mainly on the cases where the dependence of PPS ensembles on the intermediate measurements is negligibly small.
In this case, PPS ensembles are very close to {\em posterior ensembles}, a special case of PPS ensembles where no intermediate (PPS) measurements are performed.

\section{Connection states}
\label{III}

\subsection{General consideration}

Weak PPS measurements (at least, in the linear-response regime \cite{kof12}) have the important property that they do not appreciably disturb the state of the quantum system.
Hence they probe unperturbed PPS ensembles (i.e., posterior ensembles).
Let us show that weak PPS measurements are similar in a sense to conventional (preselected only) measurements.
To this end, we recall that the Born rule \rqn{7} can be rewritten equivalently as a formula for the expectation value of an observable $A$, 
 \be
\bar{A}=\sum_ia_iP_i.
 \e{72}
Namely, inserting Eq.~\rqn{7} into Eq.~\rqn{72} and using Eq.~\rqn{4}, one obtains
 \be
\bar{A}={\rm Tr}\,(A\rho).
 \e{11}
This formula describes the results of both strong and weak conventional measurements \cite{aha88,kof12}.

The weak value \rqn{2} can be recast in a form similar to the Born rule \rqn{11}, on using the invariance of the trace under cyclic permutations,
 \be
A_w={\rm Tr}\,(Aw),
 \e{9}
where \cite{hir13}
 \be
w\:=\:\frac{\rho E}{{\rm Tr}\,(\rho E)}.
 \e{10}
In the special case of pure pre- and post-selected states (i.e., $\rho=|\psi\rangle\langle\psi|$ and $E=|\phi\rangle\langle\phi|$), Eq.~\rqn{10} simplifies \footnote{In the case of pure pre- and post-selected states, unnormalized connection states were considered in Ref.~\cite{shi10}},
 \be
w=\frac{|\psi\rangle\langle\phi|}{\langle\phi|\psi\rangle}.
 \e{37}

The quantity $w$ can be called {\em connection state} (or {\em connection matrix}).
It determines the results of weak PPS measurements in the same way as the quantum state determines the results of conventional weak measurements [cf.\ Eqs.~\rqn{11} and \rqn{9}].
Therefore, the connection matrix generalizes the concept of the density matrix.
In particular, the connection matrix reduces to the density matrix of the system, $w=\rho$, when $E$ is equal or proportional to the unity operator.
This could be expected, since in this case the post-selection measurement does not yield any information or is not performed at all, and thus the posterior ensemble reduces to a preselected ensemble.

Connection states in Eq.~\rqn{10} are normalized to one, 
 \be
{\rm Tr}\,w=1.
 \e{43}
They are invariant under multiplying $\rho$ and $E$ by scalar factors, i.e., e.g., under replacing normalized $\rho$ and $E$ with unnormalized $\rho$ and $E$.
Connection states obey a time-symmetry relation, as follows.
An exchange $\rho\leftrightarrow E$, i.e., a transition from a PPS ensemble with the initial state $\rho$ and the post-selection operator $E$ to an ensemble with the (generally unnormalized) respective quantities $E$ and $\rho$, results in the change 
 \be
w\rightarrow w^\dagger.
 \e{78}
This relation implies also the change $A_w\rightarrow A_w^*$, which results on inserting Eq.~\rqn{78} into Eq.~\rqn{9} \footnote{Here we assume that $A$ is Hermitian.
Though sometimes weak values of non-Hermitian operators are discussed \cite{wis02}, here for simplicity we do not consider such weak values}.

The operators $\rho E$ and $w$ have simple physical meanings.
Indeed, we note that $\rho$ and $E$ are quantum counterparts of the prior probability distribution and the conditional probability of the measurement outcome, respectively \cite{wis10}.
This implies that $\rho E$ is a quantum counterpart of the joint probability distribution, and, hence, the connection state $w$ in Eq.~\rqn{10} is a quantum counterpart of the classical posterior probability distribution (i.e., the probability distribution conditioned by a measurement outcome); for details see Appendix \ref{B}.
In this interpretation,  Eq.~\rqn{10} is a quantum analog of Bayes' theorem [recall that ${\rm Tr}\,(\rho E)$ is the probability of the measurement outcome, cf.\ Eq.~\rqn{16}].
This shows that the weak value is a quantum analog of a classical conditional expectation value given a post-selection measurement outcome \footnote{The same conclusion was obtained from different considerations in Ref.~\cite{ste95}}.

Note, however, an important distinction of the present situation from the classical case: in quantum probability theory \cite{wis10,dre12}, $\rho$ and $E$ generally do not commute.
As a result, {\em connection matrices are generally non-Hermitian}, i.e., they have drastically different properties from conventional density matrices.

\subsection{Connection states and unusual weak values}
\label{IIIB'}

\subsubsection{The Hermitian and anti-Hermitian parts of the connection state}

The non-Hermitianity of connection states explains why weak values are generally unusual.
Indeed, the connection state can be written as a sum of the Hermitian and anti-Hermitian parts,
 \be
w=w'+iw'',\quad w'=\frac{\rho E+E\rho}{{2\rm Tr}(\rho E)},\quad
w''=\frac{[\rho,E]}{2i{\rm Tr}(\rho E)}.
 \e{12}
Here $w'$ and $w''$ are Hermitian operators, which determine the real and imaginary parts of the weak value \footnote{The Hermitian operator $w'$ was considered previously in Ref.~\cite{hof10}, where it was called ``the transient density matrix''}, 
 \be
{\rm Re}\,A_w={\rm Tr}\,(Aw'),\quad\quad{\rm Im}\,A_w={\rm Tr}\,(Aw'').
 \e{42}
Note that the condition \rqn{43} implies
 \be
{\rm Tr}\,w'=1,\quad{\rm Tr}\,w''=0.
 \e{41}
The eigenvalues of $w'$ and $w''$ can be positive or negative.
As a result, the magnitudes of these eigenvalues are not restricted by the conditions \rqn{41} and thus can be arbitrarily large.

This can be shown explicitly using the inequality
 \be
c'+c''\ge||w||,
 \e{73}
where $||w||$ is the connection-matrix norm given by the square root of the maximum eigenvalue of $w^\dagger w$ (or, equivalently, $ww^\dagger$) and $c'=||w'||$ ($c''=||w''||$) is the maximum eigenvalue magnitude for $w'$ ($w''$).
Equation \rqn{73} is implied by the property of the matrix norm, $||O_1+O_2||\le||O_1||+||O_2||$, valid for any operators $O_1$ and $O_2$ \cite{fra00}.

For example, it is easy to see that in the case of pure pre- and post-selected states, we have $w^\dagger w=|\phi\rangle\langle\phi|/|\langle\phi|\psi\rangle|^2$, yielding
 \be
||w||=|\langle\phi|\psi\rangle|^{-1}.
 \e{44}
This quantity tends to infinity for $\langle\phi|\psi\rangle\rightarrow0$.
Hence, as follows from Eq.~\rqn{73}, in this case $c'$ or $c''$ or both tend to infinity.
The absence of the upper limit for connection states gives rise to unbounded weak values and amplification \cite{aha88,kof12,hos08,gor12}.

\subsubsection{The orthogonality relation}

The Hermitian and anti-Hermitian parts of the connection state have an interesting property: they are mutually orthogonal in terms of the Hilbert-Schmidt inner product of operators,
 \be
{\rm Tr}\,(w'w'')={\rm Tr}\,(w''w')=0.
 \e{74}
To show this, we note that any operators $O_1$ and $O_2$ satisfy the identity
 \bea
&&{\rm Tr}\,[(O_1O_2+O_2O_1)(O_1O_2-O_2O_1)]={\rm Tr}(O_1O_2O_1O_2\nonumber\\
&&-O_1O_2^2O_1+O_2O_1^2O_2-O_2O_1O_2O_1)=0,
 \ea{75}
where the last equality is obtained with the help of the cyclic property of the trace.

As a consequence, we obtain the statement: for any Hermitian operators $O_1$ and $O_2$, the Hermitian and anti-Hermitian parts of their product $O=O_1O_2=O'+iO''$ are orthogonal in terms of the inner product.
Indeed, taking into account that $O'=(O_1O_2+O_2O_1)/2$ and $O''=[O_1,O_2]/(2i)$, we obtain [cf.\ Eq.~\rqn{75}]
 \be
{\rm Tr}\,(O'O'')\propto{\rm Tr}\,[(O_1O_2+O_2O_1)(O_1O_2-O_2O_1)]=0.
 \e{77}
Finally, Eq.~\rqn{74} results as a special case of Eq.~\rqn{77}.
The relation \rqn{74} shows that the Hermitian and anti-Hermitian parts of the connection state are not completely independent of each other.

\subsection{Usual and unusual connection states}
\label{IIIC}

Connection states can be classified into usual and unusual, depending on whether they allow for unusual weak values.
{\em A connection state, Eq.~\rqn{10}, is usual, if and only if $\rho$ and $E$ commute.}
Indeed, when $\rho$ and $E$ commute, $w$ is a Hermitian, positive operator, similar to a density matrix, and, as a result, any weak value is usual, i.e., a real number inside the range of the eigenvalues of the observable, just as the expectation value [cf.\ Eqs.~\rqn{11} and \rqn{9}].
On the other hand, when $\rho$ and $E$ do not commute, the connection state is unusual, i.e., then there always exists an observable with an unusual weak value.
Indeed, then $w$ is non-Hermitian and hence $w''\ne0$.
In this case, there exists a Hermitian operator $A$ such that ${\rm Tr}\,(Aw'')={\rm Im}\,A_w\ne0$; hence $A$ possesses a complex (i.e., unusual) weak value.
In particular, for pure pre- and post-selected states $|\psi\rangle$ and $|\phi\rangle$, connection states are always unusual, except for the trivial case $|\psi\rangle=|\phi\rangle$ where the posterior and preselected ensembles coincide.

Even when a connection state is unusual (i.e., non-Hermitian or non-positive), there exist observables with usual weak values.
In particular, for an arbitrary connection state, weak values are always usual whenever $A$ commutes with either $\rho$ or $E$ \cite{kof12},
 \be
[A,\rho]=0\quad{\rm or}\quad[A,E]=0.
 \e{20}
This fact can be generalized to measurements of arbitrary strength, as discussed in Sec.~\ref{IV}.

The fact that Eq.~\rqn{9} formally coincides with the Born rule implies that {\em weak PPS measurements simulate conventional measurements performed on a system in a ``quantum state''} $w$.

\subsection{Sum rules for connection states}

Consider the POVM $\{E_l\}$ for a general measurement and the corresponding connection states $w_l=\rho E_l/P_l$, where $P_l$ is the probability of the outcome $l$, Eq.~\rqn{16}.
It is easy to see that the normalization condition \rqn{71} implies the following sum rule for the connection states \cite{hir13},
 \be
\sum_lP_lw_l=\rho.
 \e{17}
Taking into account that $P_lw_l=\rho E_l$ is a quantum counterpart of a joint probability, as discussed above, we can interpret Eq.~\rqn{17} as a quantum analog of a classical marginal distribution obtained by averaging over information about the measurement.
Equation \rqn{17} is equivalent to the following sum rules,
 \be
\sum_lP_lw_l'=\rho,\quad\sum_lP_lw_l''=0,
 \e{40}
where $w_l'$ and $w_l''$ are Hermitian operators such that $w_l=w_l'+iw_l''$.
The first equality in Eq.~\rqn{40} was obtained previously in Ref.~\cite{hof10}.

An immediate consequence of Eq.~\rqn{17} is a sum rule for the weak values  $A_{w,l}={\rm Tr}\,(Aw_l)$ of a quantity $A$.
Indeed, multiplying both sides of Eq.~\rqn{17} by $A$ and taking the trace yields the sum rule
\be
\sum_lP_lA_{w,l}=\bar{A}.
 \e{29}
Taking into account that $A_{w,l}$ is a quantum counterpart of a conditional expectation, as mentioned above, Eq.~\rqn{29} is a quantum analog of the expression for an average of a conditional expectation over the measurement results.
The real and imaginary parts of Eq.~\rqn{29} provide the following sum rules,
\be
\sum_lP_l\,{\rm Re}\,A_{w,l}=\bar{A},\quad\ \ \sum_lP_l\,{\rm Im}\,A_{w,l}=0.
 \e{65}
Special cases of the sum rules \rqn{29} and \rqn{65} were obtained in Refs.~\cite{ste95,aha05a,kof12}.

\subsection{Tomography of connection states}
\label{IIIB}

Connection states can be determined experimentally.
In particular, quantum tomography of connection states can be performed with the help of weak PPS measurements, similarly to tomography of quantum states \cite{nie00,hof10}, as follows.
A connection state can be written in the form
\be 
w=\sum_{i=1}^{d^2}\alpha_iB_i, 
 \e{30}
where $\{B_i\}$ is a set of linearly-independent operators, $d$ is the dimension of the Hilbert space of the quantum system, and $\alpha_i$ are complex coefficients.
The coefficients $\alpha_i$ can be obtained on measuring a set of $d^2$ linearly-independent operators $\{A_i\}$, which may or may not coincide with $\{B_i\}$.

Indeed, on multiplying Eq.~\rqn{30} by $A_j$, taking the trace, and using Eq.~\rqn{9}, we obtain the equations,  
\be 
\sum_{i=1}^{d^2}a_{ji}\alpha_i=(A_j)_w, 
 \e{31}
where the quantities $a_{ji}={\rm Tr}\,(A_jB_i)$ are known and $(A_j)_w$ are weak values that can be measured.
Solving these equations yields $\alpha_i$ and hence the connection state $w$.
Namely, 
\be 
\alpha_i=\sum_{j=1}^{d^2}(a^{-1})_{ij}(A_j)_w, 
 \e{32}
where $a$ is the matrix with the elements $a_{ji}$.
Equations \rqn{31} and \rqn{32} simplify when $A_i$ and $B_i$ are Hermitian, since then $a_{ji}$ are real.

\section{Connection-state formalism for some measurements of arbitrary strength}
\label{IV}

\subsection{General results}
\label{IVA'}

Connection states generally cannot provide the results of strong PPS measurements.
Indeed, generally Eq.~\rqn{8} cannot be reduced to Eq.~\rqn{9}.
The reason for this is that strong measurements significantly disturb the state of the system.
However, for a special class of observables, namely for observables $A$ commuting with either $\rho$ or $E$, Eq.~\rqn{20}, there is an extension of the result \rqn{9} to PPS measurements of arbitrary strength.
Namely, in the case \rqn{20}, a PPS measurement of arbitrary strength is equivalent to (i.e., yields the same results as) the conventional (preselected only) measurement of the same strength with the effective initial state $\rho_{\rm eff}$ equal to $w$ or $w'$ \footnote{This statement with $\rho_{\rm eff}=w'$ was proved in Ref.~\cite{kof12}, Sec.~14.2.
The case $\rho_{\rm eff}=w$ is proved similarly, with the difference that now in Eq.~(14.6) in Ref.~\cite{kof12} the relation (D.7), and not (D.8), should be used}.
The equivalence of the above two values of $\rho_{\rm eff}$ (i.e., $w$ and $w'$) means that in the case \rqn{20} $w''$ does not influence the measurements, and hence one has the freedom to neglect $w''$ in $w=w'+iw''$.

Curiously, as mentioned above, $w'$ can have negative eigenvalues and hence may not represent a real quantum state.
Thus, {\em PPS measurements of any strength can simulate conventional measurements of observables in a fictitious state described by a ``density matrix'' with negative eigenvalues}.

In particular, strong PPS measurements can simulate ideal projective measurements involving the ``quantum state'' $w$ or $w'$.
This can be derived directly from Eq.~\rqn{8}.
Namely, on taking into account Eq.~\rqn{23}, the cyclic property of the trace, and the fact that when $[A,\rho]=0\ ([A,E]=0)$, then $\Pi_i$ commutes with $\rho\ (E)$, Eq.~\rqn{8} can be transformed to the form
 \be
P_{i|E}={\rm Tr}\,(\Pi_iw)\equiv\Pi_{i,w}, 
 \e{21}
where $\Pi_{i,w}$ is the weak value of $\Pi_i$.
Similarly, we obtain that $P_{i|E}={\rm Tr}\,(\Pi_iw^\dagger)$, yielding, in view of Eq.~\rqn{21}, $P_{i|E}={\rm Tr}\,[\Pi_i(w+w^\dagger)/2]={\rm Tr}\,(\Pi_iw')$.
Thus we obtain a result alternative to Eq.~\rqn{21},
 \be
P_{i|E}={\rm Tr}\,(\Pi_iw'). 
 \e{36}
Equations \rqn{21} and \rqn{36} coincide with the result of a conventional strong measurement in Eq.~\rqn{7} with $\rho$ substituted by $w$ or $w'$.
Note that the probabilities in Eqs.~\rqn{21} and \rqn{36} are classical, i.e., non-negative, although generally $w$ is non-Hermitian and $w'$ is not positive; indeed, the probabilities in Eq.~\rqn{8} are easily seen to be non-negative.

More generally, as shown in Appendix \ref{A}, in the case \rqn{20} PPS measurements of arbitrary strength provide the weak value $A_w$. Now $A_w$ is a usual value of $A$, i.e., a real value within the range of the eigenvalues of $A$.

\subsection{Violation of an uncertainty relation in connection states}
\label{IVA}

It is interesting to note that though in the case \rqn{20} PPS measurements are usual from the point of view of classical physics, they are not usual from the point of view of quantum mechanics.
Indeed, Eq.~\rqn{21} implies the well known fact that in a PPS ensemble with pure pre- and post-selected states $|\psi\rangle$ and $|\phi\rangle$, any observable with an eigenstate $|\psi\rangle$ or $|\phi\rangle$ has a definite value, equal to the corresponding eigenvalue \cite{aha90}.
Therefore in such an ensemble two non-commuting observables without common eigenvectors (such as two orthogonal projections of the angular momentum) can have definite values.
In contrast, such a situation is not possible in a conventional (preselected only) ensemble.
In view of the above discussion, this ``paradox'' \cite{aha90} can be interpreted as a direct consequence of the unusual character of the connection matrix, since in the present case arbitrary-strength PPS measurements simulate conventional measurements of a system in the ``quantum state'' $w$ or $w'$.

Consider an extension of the above unusual situation.
In the case when $\rho$ and $E$ are not pure states, one can expect that the uncertainties of two non-commuting operators which commute with $\rho$ and $E$, respectively, can be still below those required by quantum mechanics for conventional measurements.
We show this for the special case of a qubit, where the following sum uncertainty relation holds \cite{hof03},
 \be
\Delta\sigma_1^2+\Delta\sigma_2^2\ge1.
 \e{46}
Here $\Delta\sigma_i^2$ is the variance of the Pauli matrix $\sigma_i=\Pi_{i+}-\Pi_{i-}$, where $\Pi_{i\pm}={|i\pm\rangle\langle i\pm|}$.

Consider arbitrary-strength PPS measurements such that $\rho$ ($E$) commutes with  $\sigma_1$ ($\sigma_2$), 
 \bea
&&\rho=p_+\Pi_{1+}+p_-\Pi_{1-}=(I+\lambda_1\sigma_1)/2,\nonumber\\ 
&&E=e_+\Pi_{2+}+e_-\Pi_{2-}=[(e_++e_-)/2](I+\lambda_2\sigma_2).\
 \ea{49}
Here $\lambda_1=p_+-p_-$ and $\lambda_2=(e_+-e_-)/(e_++e_-)$, so that $-1\le\lambda_{1,2}\le1$.
The parameters $|\lambda_1|$ and $|\lambda_2|$ are measures of purity of $\rho$ and $E$, so that, e.g., for $|\lambda_1|=1$ $\rho$ is pure and for $\lambda_1=0$ $\rho$ is completely mixed ($\rho\propto I$).
It is easy to show that Eq.~\rqn{10} yields now $w=w'+iw''$ with
\bes{67'}
 \bea
&&w'=(I+\lambda_1\sigma_1+\lambda_2\sigma_2)/2,\label{50}\\
&&w''=\lambda_1\lambda_2\sigma_3/2.
 \ea{67}\ese
Note that here $w'$ and $w''$ satisfy the orthogonality relation \rqn{74}.

In the present case, the variance of $\sigma_i$ is $\Delta\sigma_{i,w}^2=(\sigma_i^2)_w-(\sigma_i)_w^2$.
Taking into account that $(\sigma_i^2)_w=(I)_w=1$ and
 \be
(\sigma_i)_w={\rm Tr}\,(\sigma_iw')=\lambda_i\quad(i=1,2), 
 \e{66}
we obtain that $\Delta\sigma_{i,w}^2=1-\lambda_i^2$.
These variances are usual values, $0\le\Delta\sigma_{i,w}^2\le1$, as one should expect.
The sum of the variances is
 \be
\Delta\sigma_{1,w}^2+\Delta\sigma_{2,w}^2 =2-\lambda_1^2-\lambda_2^2.
 \e{47}
Note that here the purity parameters for $\rho$ and $E$ enter on an equal footing.

\begin{figure}[htb]
\begin{center}
\includegraphics[width=7cm]{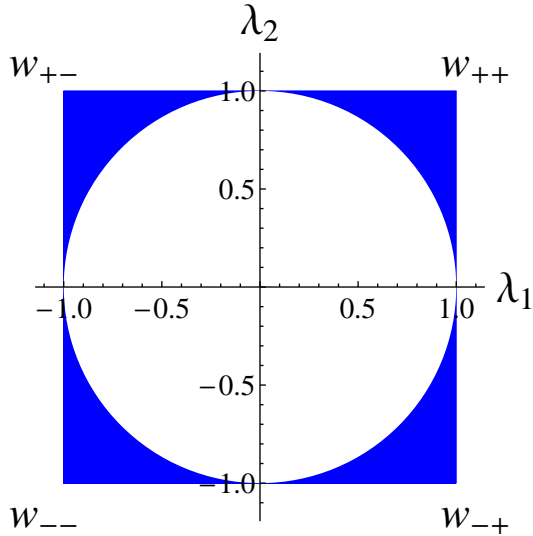}
\end{center}
\caption{The square $|\lambda_{1,2}|\le1$ provides the allowed region of the parameters.
The connection states \rqn{79} with definite values of $\sigma_1$ and $\sigma_2$ correspond to the vertices of the square ($\lambda_{1,2}=\pm1$), as shown.
The shaded part of the parameter space shows the region \rqn{48} where the uncertainty relation is violated, Eq.~\rqn{51}, and $w'$ is unusual (non-positive).
}
 \label{f2}\end{figure}

Consider special cases of Eq.~\rqn{47}.
For pure pre- and post-selected states, we have $|\lambda_1|=|\lambda_2|=1$, and Eq.~\rqn{47} implies that $\Delta\sigma_{1,w}=\Delta\sigma_{2,w}=0$, i.e., $\sigma_1$ and $\sigma_2$ have definite values, in agreement with the above discussion.
The connection states for this case can be expressed by Eq.~\rqn{37} in the form (see also Fig.~\ref{f2})
 \be
w_{\alpha\beta}=(1+i\alpha\beta)|2\alpha\rangle\langle1\beta|\quad\ (\alpha,\beta=\pm).
 \e{79}
Furthermore,  when
 \be
\lambda_1^2+\lambda_2^2>1,
 \e{48}
we have 
 \be
\Delta\sigma_{1,w}^2+\Delta\sigma_{2,w}^2<1, 
 \e{51}
in contradiction to the uncertainty relation \rqn{46}.
The condition \rqn{48}, shown graphically in Fig.~\ref{f2}, also ensures that $w'$ in Eq.~\rqn{50} is unusual, i.e., has a negative eigenvalue, since the eigenvalues of $w'$ are $(1\pm\sqrt{\lambda_1^2+\lambda_2^2})/2$.
The left-hand side of Eq.~\rqn{47} increases with the decrease of the purity of $\rho$ or $E$ or both.
Finally, when $\lambda_1^2+\lambda_2^2\le1$, the quantity $w'$ is usual, i.e., a positive operator. 
In this case, we obtain that $\Delta\sigma_{1,w}^2+\Delta\sigma_{2,w}^2\ge1$, in agreement with Eq.~\rqn{46}, as one should expect.

Thus, we have shown that the uncertainty relation \rqn{46} is violated in PPS measurements whenever $w'$ is unusual, i.e., in the case \rqn{48}.
Now the value of $w''$ is unimportant for measurements of $\sigma_1$ and $\sigma_2$, as discussed in Sec.~\ref{IVA'}.
Note, however, that in the present case the connection state $w$ is unusual for almost all values of the parameters, since $w''$ is nonzero unless $\rho$ or $E$ is a multiple of $I$ (i.e., unless $\lambda_1$ or $\lambda_2$ vanishes); cf.\ Eq.~\rqn{67}.
In other words, $w$ is unusual in the whole square in Fig.~\ref{f2}, except for the axes lines.
This agrees with the statement in Sec.~\ref{IIIC} that $w$ is unusual whenever $\rho$ and $E$ do not commute.
Generally the non-Hermitian part $w''$ of the connection state is important for weak measurements, since $w''$ is responsible for complex weak values [cf.\ Eq.~\rqn{42}].
For example, now $(\sigma_3)_w={\rm Tr}\,(\sigma_3w)=i{\rm Tr}\,(\sigma_3w'')=i\lambda_1\lambda_2$ [cf.\ Eqs.~\rqn{67'}].

\subsection{Retrodictive states}

An important special case of the present theory occurs when the initial state is not known, i.e., it is completely random, $\rho\propto I$.
Measurements in this case are called post-selected only measurements \cite{aha02}.
Now the connection state \rqn{10} becomes
 \be
w\:=\:\frac{E}{{\rm Tr}\,E}\equiv\rho_{\rm retr},
 \e{45}
this operator being called the retrodictive state \cite{bar00}.
Note that the retrodictive state describes post-selected only measurements of {\em any strength} for {\em any observables}, since any operator $A$ commutes with $\rho\propto I$ [cf.\ the first equality in Eq.~\rqn{20}].
The study of the retrodictive state \rqn{45} provides information on the detector \cite{amr11}.

\subsubsection{Tomography of quantum detectors}

One can perform tomography of the retrodictive state and thus reconstruct $E$ by the method described in Sec.~\ref{IIIB}.
The only difference is that now measurements of any strength can be used.
This method differs from the usual tomography of quantum detectors \cite{lun09a} which employs conventional (preselected only) measurements of the probabilities $P_i={\rm Tr}\,(\rho_i\,E)$ for a set of different initial states $\{\rho_i\}$.
In the present method the initial state is completely random, and the (post-selected only) measurements are performed for a set of linearly-independent operators $\{A_i\}$.

The above method provides $\rho_{\rm retr}$, which determines the POVM operator $E=({\rm Tr}\,E)\rho_{\rm retr}$ only with the accuracy to the factor ${\rm Tr}\,E$.
This factor results easily from a measurement of the post-selection probability $P={\rm Tr}\,(\rho E)$.
Indeed, since now $\rho=I/d$, we have $P={\rm Tr}\,E/d$, and hence ${\rm Tr}\,E=Pd$.

\subsubsection{Symmetric form of connection states}

Using the definition of retrodictive states in Eq.~\rqn{45}, the expression \rqn{45} for the connection states can be recast in a symmetric form,
 \be
w\:=\:\frac{\rho_{\rm pred}\rho_{\rm retr}}{{\rm Tr}\,(\rho_{\rm pred}\rho_{\rm retr})},
 \e{68}
where the predictive state $\rho_{\rm pred}=\rho$.
This expression shows explicitly that quantum smoothing is a combination of prediction provided by the quantum state $\rho$ and retrodiction determined by the retrodictive state $\rho_{\rm retr}$.

\section{Connection states and quantum complementarity}
\label{V}

In this section, we show that the unusual character of the connection states in the generic case where $\rho$ and $E$ do not commute is a manifestation of the non-classical nature of quantum mechanics or, more specifically, quantum complementarity.
Indeed, the complementarity principle states, loosely speaking, that different measurements generally provide results contradicting each other from the point of view of usual (classical) logic, since measurements of non-commuting observables provide information on {\em incompatible aspects} of a quantum system, such as, e.g., wave-like and particle-like behavior.
As a result, one cannot use information provided by a measurement of some observable to improve one's knowledge on a quantum system obtained from a preceding measurement of a different non-commuting observable, since the two pieces of information describe incompatible aspects of a quantum system.

The initial state $\rho$ provides the maximum information on observables that commute with $\rho$ and thus describes a certain aspect of the quantum system represented by such an observable.
(If $\rho$ has degenerate eigenvalues, the observables commuting with $\rho$ may not commute with each other; then $\rho$ describes two or more incompatible aspects of the quantum system.)
When $E$ commutes with $\rho$, the measurement is in essence classical, and it can improve our knowledge on the system.
This improved knowledge is described by the connection state $w$ that is now a usual state (a positive operator).
Correspondingly, now Eq.~\rqn{10} coincides in essence with the classical Bayes theorem.

In contrast, when $E$ and $\rho$ do not commute, the final measurement probes an aspect of the quantum system which is {\em incompatible} with the aspect(s) described by the initial state $\rho$.
Since now information provided by the measurement cannot increase the knowledge described by $\rho$, the connection state $w$ must be necessarily unusual (i.e., a non-Hermitian or non-positive operator).
Otherwise, $w$ would provide an improved knowledge on the quantum system, and this is forbidden by the complementarity principle.
From this we can deduct that unusual weak values have no direct physical meaning in the classical sense.
An unusual value simply indicates that the property of the quantum system probed by a given PPS measurement is non-classical, i.e., it cannot be described in classical terms.

Of course, this does not mean that weak values are useless.
On the contrary, they have unique properties, such as amplification, which are very useful, as mentioned above.
Moreover, they provide a more direct access to quantum information than conventional measurements, as implied by the above discussion and demonstrated in experiments on direct measurements of the wave function \cite{lun11,*sal13,koc11,*bli13}.

In summary, the above discussion shows that the reason for the unusual connection states is the non-classical nature of quantum mechanics or, more specifically, quantum complementarity.
The latter is also the ultimate reason for counterintuitive results of weak and some non-weak PPS measurements, including unusual weak values, quantum paradoxes \cite{aha05}, and violations of uncertainty relations (Sec.~\ref{IVA}), since such results can be explained as resulting from unusual connection states.

\section{Time dependence}
\label{VI}

\subsection{Time dependence of general PPS measurements}

Until now, we neglected the effects of the free dynamics of the quantum system on PPS measurements.
It is instructive to include into the consideration the time dependence due to the system Hamiltonian $H(t)$.
We assume that the initial state is prepared at time $t_0$ and the post-selection measurement is performed at time $t_1>t_0$.
Moreover, we assume that a PPS measurement is performed using the von-Neumann scheme \cite{neu55,aha88}, so that the system and meter are correlated impulsively at time $t\ (t_0<t<t_1)$.
Then the system dynamics is taken into account most generally by the substitutions in the formulas for PPS measurements [see, e.g., Eqs.~\rqn{8} and \rqn{2}] given by Eqs.~\rqn{C3'} in Appendix \ref{C}.

\subsubsection{Schr\"{o}dinger picture}

In particular, it is common to describe the time dependence of PPS measurements by the substitutions \cite{aha90,aha91,aha05} [see Eqs.~\rqn{C3'} with $U_1=I$]
\bes{57'}
 \bea
&&\rho\:\rightarrow\:\rho(t)=U(t,t_0)\rho\, U^\dagger(t,t_0),\label{57}\\
&&E\:\rightarrow\: E(t_1,t)=U^\dagger(t_1,t)E\,U(t_1,t).
 \ea{33}\ese
Here 
 \be
U(t',t'')=T\exp\left[-\frac{i}{\hbar}\int_{t''}^{t'}d\tau H(\tau)\right], 
 \e{58}
where $T$ is the chronological operator.
In Eq.~\rqn{33}, $E(t_1,t)$ is the POVM operator in the Heisenberg picture defined with respect to the initial time $t$.
As a function of $t$, $E(t_1,t)$ satisfies the backward Heisenberg equation \cite{bal93}, 
 \be
i\hbar\frac{dE}{dt}=[H(t),E],
 \e{52}
with the ``final'' condition $E(t_1,t_1)=E$.
In the representation \rqn{57'}, the observable $A$ is independent of the system dynamics.
Therefore, this representation can be called ``the Schr\"{o}dinger picture for PPS measurements''.

\subsubsection{Heisenberg pictures}

More generally, PPS measurements can be described using the Heisenberg picture with respect to the arbitrary reference time $t_r\ (t_0\le t_r\le t_1)$,  
\bes{59'}
 \bea
&&A\:\rightarrow\: A(t,t_r)=U^\dagger(t,t_r)A\,U(t,t_r),\label{59}\\
&&\rho\:\rightarrow\:\rho(t_r),\label{60}\\
&&E\:\rightarrow\: E(t_1,t_r).
 \ea{61}\ese
This representation results from Eqs.~\rqn{C3'} with $U_1=U^\dagger(t,t_r)$.
A special case of Eqs.~\rqn{59'} for $t_r=t$ is the common representation \rqn{57'}.
Moreover, one can consider the forward and backward Heisenberg pictures, respectively, 
 \be
A\rightarrow A(t,t_0),\quad E\rightarrow E(t_1,t_0)
 \e{62}
and
 \be
A\rightarrow A(t,t_1),\quad\rho\rightarrow\rho(t_1).
 \e{63}
obtained from Eqs.~\rqn{59'} for $t_r=t_0$ and $t_r=t_1$, respectively.
In the Heisenberg pictures, Eqs.~\rqn{62} and \rqn{63}, $\rho$ and $E$ are fixed at either the initial or the final time, and only the observable evolves: forward in time from $t_0$ to $t$ in Eq.~\rqn{62} or backward in time from $t_1$ to $t$ in Eq.~\rqn{63}.

The above representations, Eqs.~\rqn{57'} and \rqn{59'}-\rqn{63}, simplify when one of the quantities $A,\ \rho$, or $E$ commutes with the Hamiltonian $H(t)$.
For example, it is interesting that, when $A$ is a constant of motion [i.e., commutes with $H(t)$], the system dynamics in PPS measurements can be described simply by replacing the initial state $\rho$ by its final value $\rho(t_1)$ [cf.\ Eq.~\rqn{63}].

\subsection{Time dependence of connection states: Time-dependent quantum Bayes formula}


Inserting the substitutions \rqn{57'} into Eq.~\rqn{10} yields the time-dependent connection matrix, 
 \be
w(t)=\rho(t)E(t_1,t)/P,
 \e{18}
where 
 \be
P={\rm Tr}\,[\rho(t) E(t_1,t)]={\rm Tr}\,[\rho E(t_1,t_0)]={\rm Tr}\,[\rho(t_1) E].
 \e{53}
Note that the normalization factor $P$ is independent of time, being equal just to the probability of the measurement outcome described by $E$, as indicated by the last expression in Eq.~\rqn{53}.

Equation \rqn{18} can be interpreted in terms of probabilities.
The operator $E(t_1,t)$ is a quantum counterpart of the probability of a measurement outcome at $t_1$ given a system state at $t$; correspondingly, $\rho(t)E(t_1,t)$ is a quantum counterpart of the joint probability of a system state at $t$ and a measurement outcome at $t_1$.
This interpretation agrees with the above fact that ${\rm Tr}\,[\rho(t) E(t_1,t)]$ is the (unconditional) probability of the measurement outcome.
Thus, the time dependence of $w(t)$ is formally the same as for a classical posterior probability distribution.
Hence, Eq.~\rqn{18} can be called {\em the time-dependent quantum Bayes formula}.

A more explicit form of Eq.~\rqn{18} is
 \be
w(t)\:=\:U(t,t_0)\rho\, U^\dagger(t_1,t_0)E\,U(t_1,t)/P.
 \e{13}
In the case of pure pre- and post-selected states, Eq.~\rqn{13} becomes
 \be
w(t)=\frac{U(t,t_0)|\psi\rangle\langle\phi|U(t_1,t)}{\langle\phi|U(t_1,t_0)|\psi\rangle}.
 \e{34}
Connection states obey the von Neumann equation for time evolution,
 \be
i\hbar\frac{dw}{dt}=[H(t),w].
 \e{14}
One can solve Eq.~\rqn{14} either forward in time using the initial condition $w(t_0)=\rho E(t_1,t_0)/P$ or backward in time using the final condition $w(t_1)=\rho(t_1)E/P$.

\subsubsection{Weak values}

The weak value in the presence of the free evolution of the quantum system is given by 
 \be
A_w={\rm Tr}\,[Aw(t)]={\rm Tr}\,[A\rho(t)E(t_1,t)]/P.
 \e{15}
In addition to this ``Schr\"{o}dinger picture'', one can use also various Heisenberg pictures, Eqs.~\rqn{59'}-\rqn{63}.
In particular, Eq.~\rqn{15} can be recast in the ``forward Heisenberg picture'', 
 \be
A_w={\rm Tr}\,[A(t,t_0)w(t_0)], 
 \e{55}
or in the ``backward Heisenberg picture'', 
 \be
A_w={\rm Tr}\,[A(t,t_1)w(t_1)].
 \e{56}
In the Heisenberg pictures, Eqs.~\rqn{55} and \rqn{56}, the connection matrix is constant in time, being equal to its initial or final value, $w(t_0)$ or $w(t_1)$, respectively.

\subsubsection{Retrodictive states and the symmetric form for $w(t)$}

When $\rho\propto I$ [and hence also $\rho(t)\propto I$], then Eq.~\rqn{18} yields a time-dependent retrodictive state
 \be
\rho_{\rm retr}(t)=\frac{E(t_1,t)}{{\rm Tr}\,E}.
 \e{69}
Retrodictive states satisfy Eq.~\rqn{52} or \rqn{14} with the final condition $\rho_{\rm retr}(t_1)=E/{\rm Tr}\,E$.
Retrodictive states differ essentially from usual (predictive) quantum states, since the former (the latter) describe post-selected (preselected) ensembles.
However, there is a remarkable symmetry between them: retrodictive states $\rho_{\rm retr}(t)$ can be interpreted as usual states propagating backward in time, from $t_1$ to $t$ \cite{aha91,bar00,peg08}.
With the help of Eq.~\rqn{69}, $w(t)$ in Eq.~\rqn{18} can be recast in a symmetric form,
 \be
w(t)=\frac{\rho_{\rm pred}(t)\rho_{\rm retr}(t)}{P'},
 \e{70}
where $\rho_{\rm pred}(t)=\rho(t)$ and $P'={\rm Tr}\,[\rho_{\rm pred}(t)\rho_{\rm retr}(t)]=P/{\rm Tr}\,E\ (0<P'\le1)$.
Equation \rqn{70} is a time-dependent analog of Eq.~\rqn{68}.

\subsection{Some remarks}

Conventional measurements are local in time, since their results depend only on the values of the density matrix and/or the observable at the moment of measurement.
In contrast, PPS measurements are non-local in time, since they combine a measurement of $A$ at $t$ and the post-selection measurement at $t_1$.
Thus, the time dependence in Eqs.~\rqn{57'} and hence in Eq.~\rqn{18} can be explained as follows.
Since the measurement of $A$ is made at $t$, the measurement result should depend on the value of the density matrix at $t$, which explains Eq.~\rqn{57} and the first factor in Eq.~\rqn{18}.
Next, the post-selection measurement is performed after the measurement of $A$, at $t_1\ge t$.
Since the state of the system changes with time, the probability of the post-selection outcome and hence the PPS ensemble and its statistical properties may vary with time.
This change is compensated for by employing the POVM operator $E$ in the Heisenberg picture in the interval $(t,t_1)$, which results in a time-independent post-selection probability; cf.\ Eq.~\rqn{53}.
This explains the time dependence in Eq.~\rqn{33} and in the second factor in Eq.~\rqn{18}.

Let us remark on the measurement probability $P$ in Eq.~\rqn{53}.
This is the usual Born rule, written in different representations.
Though the differences between the meanings of different expressions in Eq.~\rqn{53} were discussed \cite{bar99,bar00,peg08} at some length \footnote{In particular, in Refs.~\cite{bar99,bar00} the first two expressions in Eq.~\rqn{53} are called the ``retrodictive formalism'' in quantum mechanics and the third expression is called the ``predictive formalism'', whereas in Ref.~\cite{peg08} a relation to causality in quantum mechanics was considered}, the operator products under the trace sign in Eq.~\rqn{53} cannot be observed in conventional measurements.
The present theory implies that not only the trace in Eq.~\rqn{53} but also the quantity under the trace sign are experimentally accessible. 
Since this quantity is actually an unnormalized connection state, it can be probed and tomographically reconstructed with the help of weak PPS measurements, as described above.

\subsubsection{Evolution backward in time}

Equation \rqn{18} shows that the state of the system between the preparation and measurement, described by the connection matrix $w(t)$, is determined not only by the earlier preparation event but also by the later measurement event.
This counterintuitive fact can be viewed as resulting from another counterintuitive fact, namely, that the retrodictive state evolves backward in time \cite{aha91,bar99,bar00,aha05,aha02,hof10,shi10,aha10,aha09}.

The above interpretations can be useful in applications since they provide a clear (though unusual) physical picture.
However, they should not be taken too literally, since they are not unique, owing to the fact that the common representation \rqn{57'}, for which Eq.~\rqn{18} holds, is not unique.
In particular, note that, at least, in one of the representations, the backward Heisenberg picture [Eqs.~\rqn{63} and \rqn{56}], $E$ does not depend on time and hence cannot be interpreted as a backward evolving state.
[Instead, in this case the observable $A(t,t_1)$ is moving backward in time.]
Thus, generally, there is a freedom of choice of the representation and/or the interpretation in PPS measurements.
This freedom allows one to choose the representation/interpretation which is most suitable for a given problem; see examples in Refs.~\cite{bar99,bar00,peg08}.

\subsubsection{Retrospective nature of connection states}

From the point of view of orthodox quantum mechanics, connection states have a retrospective nature.
Indeed, before the final measurement a PPS ensemble does not exist, since the property probed by the final measurement does not exist, even in a hidden form, at least, when $E$ and $\rho$ do not commute.
As a result, before the final measurement, the connection state is a purely theoretical entity without a real physical counterpart and hence without a physical meaning.
After the creation of a PPS ensemble, the connection state acquires a physical meaning retroactively.
However, the description provided by connection states is retrospective, since it is possible only at times later than the temporal interval where it is applicable.

Here we encounter one of paradoxical features of quantum mechanics.
Namely, a quantum measurement not only reveals a property of a quantum system, it also creates the whole history of the system from the preparation to the measurement.
(Here by the ``history'' we mean the description of the behavior of the quantum system with the help of a connection state.)
In particular, if an experimenter would decide to change the measurement setting and thus change the set of the POVM operators $\{E_l\}$, this would provide a different set of the posterior ensembles and hence quite different histories.
In view of the retrospective nature of connection states, the backward-in-time propagation effects described by them do not violate the principle of causality \cite{peg08}.

\section{Discussion}
\label{VII}

\subsection{Comparison to the two-state vector formalism}

Let us compare the present connection-state formalism (CSF) with the two-state vector formalism (TSVF) developed by Aharonov et al.\ \cite{aha64,aha02,aha10}.
(Two-state vectors are called also two-time states \cite{aha09}.)
The TSVF is applicable to weak and strong (but not intermediate-strength) PPS measurements, whereas the CSF holds for weak and some arbitrary-strength PPS measurements.
Moreover, the TSVF describes the special case of pure pre- and post-selected states, whereas our formalism describes the general case with arbitrary initial states and post-selection measurements.
Thus, for many cases only one of the two formalisms is applicable, whereas both formalisms hold for weak and some strong PPS measurements with pure pre- and post-selected states.

A two-state vector has a rather unusual mathematical structure: it is ``a mathematical object which is comprised of a bra and a ket vector with an empty slot in between'' \cite{aha09}.
In contrast, a connection matrix is a familiar and very well studied mathematical object, similar to other objects in the quantum formalism, namely, a linear operator on the Hilbert space of the quantum system.
A relative simplicity of the CSF allowed us to derive results, which cannot be obtained in the frame of the TSVF, such as those related to the separation of connection states into the Hermitian and anti-Hermitian parts (cf.\ Sec.~\ref{IIIB'}).

Another advantage of the CSF is that it describes PPS measurements in a manner similar to the Born rule [see Eq.~\rqn{9}], thus allowing one to study PPS measurements by analogy with conventional measurements.
Note that in the case of pure pre- and post-selection states, there is a simple formal relation between the TSVF and the CSF: the connection matrix is given by the normalized direct product of the two vectors comprising the two-state vector, see Eq.~\rqn{37}.

\subsection{Connection and posterior states}

The above analysis focused on one quantity characterizing a posterior ensemble: the connection state $w$ used for quantum smoothing in the past ($t_0<t<t_1$).
There is also another quantity characterizing a posterior ensemble:  the conventional posterior quantum state $\rho'$ used for predictions in the future ($t>t_1$).
Let us compare $w$ and $\rho'$.

In contrast to $w$, the posterior state $\rho'$ is not defined uniquely for given $\rho$ and $E$.
We restrict the consideration to the important case when the post-selection measurement is minimally disturbing, so that \cite{wis10}
 \be
\rho'\:=\:\frac{\sqrt{E}\,\rho\,\sqrt{E}}{{\rm Tr}\,(\rho E)}.
 \e{19}
Similarly to Eq.~\rqn{10}, the equality \rqn{19} can be also considered as a quantum counterpart of Bayes' theorem \cite{kor99,gar04}.
However, there is a significant difference between the two above quantities.
Both expressions involve information on two (generally) incompatible aspects of a quantum system provided by $\rho$ and $E$.
However, in the connection state, Eq.~\rqn{10}, the two pieces of information enter on an equal footing.
As a result, PPS measurements yield generally unusual results, as discussed in Sec.~\ref{V}.

In contrast, $\rho'$ is a conventional state which describes classical results of conventional measurements.
This reconciliation of the two incompatible aspects is achieved by a complete or partial projection of the initial state performed in Eq.~\rqn{19}, which erases (completely or partially) the information contained in the initial state $\rho$.
The degree of erasure increases with the measurement strength, the complete erasure being achieved for projective measurements with rank-1 projectors [cf.\ the remark after Eq.~\rqn{35}].

It is interesting that there are situations where $\rho'=w$.
This occurs when $E$ commutes with $\rho$.
In particular, $\rho'=w=\rho_{\rm retr}$ for any $E$ when $\rho$ is completely random, $\rho\propto I$.
In other words, when the post-selection measurement is minimally disturbing, the retrodictive state \rqn{45} coincides with the posterior (i.e., predictive) state.
In this case, measurements (strong or weak or intermediate-strength) yield the same results, irrespective of whether they are performed before or after the post-selection measurement.

\section{Conclusions}
\label{VIII}

In the present paper, we consider the notion of connection state (or connection matrix) that describes quantum systems in a posterior ensemble produced by a quantum measurement.
{\em A connection matrix is a non-Hermitian operator that is a direct extension of the density matrix.}
The present formalism provides a unified description of various types of measurements, including conventional, retrodictive, as well as weak and certain non-weak PPS measurements.

We have discussed the physical meaning of connection matrices and their relation to PPS measurements and weak values.
We have shown that the unusual character of weak values is a direct consequence of the non-Hermitianity of connection matrices, which in turn results from the non-classical nature of quantum mechanics, namely, from the complementarity principle.
Next, it is shown, in essence, that weak and some arbitrary-strength PPS measurements allow one to perform quantum simulations of non-positive/non-Hermitian quantum states.
In the future, it would be of interest to consider implications of this fact to quantum information processing.

Thus, here it is shown that a broad class of non-Hermitian operators are experimentally accessible quantities with a clear physical meaning.
The present approach can be useful in all applications of PPS measurements, including quantum information processing.

\acknowledgments
We thank S. Ashhab and J. Dressel for discussions.
This work is partly supported by the ARO, RIKEN iTHES Project, MURI Center for Dynamic Magneto-Optics, JSPS-RFBR contract No. 12-02-92100, Grant-in-Aid for Scientific Research (S), MEXT Kakenhi on Quantum Cybernetics, and the JSPS via its FIRST program. 
SKO thanks Dr.\ L.\ Yang for continuous support.

\appendix
\section{Classical analogy for quantum measurements}
\label{B}

In the classical case, a physical observable $A$ can be often described as a random variable which can assume the values $\{a_i\}$ with the probabilities $\{p_i\}$ (for simplicity, we consider an observable with discrete values).
Generally, a measurement of $A$ is non-ideal; it is described by the probability $e_{i,l}$ of a certain measurement outcome $l$ provided that the value of $A$ is $a_i$.
Thus the quantities 
 \be
P_l=\sum_ip_ie_{i,l},\quad P_{i,l}=p_ie_{i,l},\quad P_{i|l}=\frac{p_ie_{i,l}}{\sum_ip_ie_{i,l}}
 \e{B1}
are, respectively, the probability of the measurement outcome $l$, the joint probability of the value $a_i$ and the outcome $l$, and the conditional probability of $a_i$ subject to the outcome $l$.
The latter equality is the (classical) Bayes theorem.

In quantum mechanics, the classical case can be simulated if the three operators $A,\ \rho$, and $E_l$ commute with each other.
Then one can find a basis $\{|i\rangle\}$, in which they are diagonal simultaneously, with the eigenvalues corresponding to the eigenvector $|i\rangle$ equaling $a_i,\ p_i$, and $e_{i,l}$, respectively.
(In this Appendix, we assume for simplicity that the eigenvalues $a_i$ of $A$ are non-degenerate.)
In this case, if we consider only the non-degenerate observables diagonal in $\{|i\rangle\}$, the quantities $p_i$ and $e_{i,l}$ have essentially the same meaning as in the classical case.
Namely, $\rho$ provides the probability distribution of the states $|i\rangle$ and, hence, of the values $a_i$, whereas the operators $E_l$ provide the conditional probability distribution of the measurement results.

Now, the quantities in Eq.~\rqn{B1} can be written equivalently in the form
 \be
P_l={\rm Tr}(\rho E_l),\quad P_{i,l}=(\rho E_l)_{ii},\quad P_{i|l}=\frac{(\rho E_l)_{ii}}{{\rm Tr}(\rho E_l)},
 \e{B2}
which implies that $\rho E_l$ and $\rho E_l/{\rm Tr}(\rho E_l)$ provide unconditional and conditional probability distributions.
In the general case, where $A,\ \rho$, and $E_l$ do not necessarily commute with each other, $\rho E_l$ is not a classical distribution [although ${\rm Tr}(\rho E_l)$ remains the probability of the outcome $l$].
Still, in a sense, $\rho E_l$ and $\rho E_l/{\rm Tr}(\rho E_l)$ can be viewed as quantum counterparts of the classical unconditional and conditional probability distributions.

\section{Measurements of arbitrary strength}
\label{A}

In the case \rqn{20}, PPS measurements of arbitrary strength provide a usual value of $A$, i.e., a real value within the range of the eigenvalues of $A$.
To show this, we note that, as follows from the above, in the case \rqn{20} we can consider a PPS measurement of $A$ as a conventional measurement performed in the effective quantum state $w$.
We use the von Neumann measurement model, where the system and the meter are correlated by a unitary transformation $U_c(A)$, which is defined in the tensor product of the Hilbert spaces of the system and the meter.
The operator $U_c(A)$ is a function of $A$ such that $U_c(a_i)\ne U_c(a_j)\ \forall i\ne j$.
A simple example of $U_c(A)$ is $U_c(A)=\exp(-iA\otimes F)$, where $F$ is a meter observable \cite{aha88,kof12}.
The application of $U_c(A)$ is followed by a measurement of the expectation value $\bar{R}_f$ of the pointer variable $R$ of the meter, yielding \cite{kof12}
 \be
\bar{R}_f={\rm Tr}\,[(I\otimes R)U_c(A)(w\otimes\rho_{\rm
M})U_c^\dagger(A)],
 \e{22}
where $\rho_{\rm M}$ is the initial state of the meter.

Assume first that $A$ commutes with $\rho$.
Then a basis $\{|ik\rangle\}$ can be found where $A$ and $\rho$ are diagonal,
 \bea
&&A=\sum_{i,k}a_i|ik\rangle\langle ik|,\label{24}\\ 
&&\rho=\sum_{i,k}p_{ik}|ik\rangle\langle ik|.
 \ea{A5}
Here the quantum number $k$ takes into account a possible degeneracy of the eigenvalues $a_i$.
Using Eqs.~\rqn{10} and \rqn{24} and the cyclic property of the trace, Eq.~\rqn{22} can be transformed to the form
 \be
\bar{R}_f=\sum_{i,k}P_{ik}\bar{R}_{f,i}.
 \e{25}
Here $\bar{R}_{f,i}$ is the pointer expectation value corresponding to the eigenvalue $a_i$,
 \be
\bar{R}_{f,i}={\rm Tr}\,[RU_c(a_i)\rho_{\rm M}U_c^\dagger(a_i)],
 \e{26}
and $\{P_{ik}\}$ is the probability distribution,
 \be
P_{ik}=\frac{p_{ik}\langle ik|E|ik\rangle}{{\rm Tr}\,(\rho E)}=\frac{p_{ik}\langle ik|E|ik\rangle}{\sum_{i,k}p_{ik}\langle ik|E|ik\rangle}.
 \e{27}
Taking into account that $\Pi_i=\sum_{k}|ik\rangle\langle ik|$, it is easy to see that 
 \be
\sum_kP_{ik}={\rm Tr}\,(\Pi_iw)=\Pi_{i,w} 
 \e{A2}
[cf.\ Eq.~\rqn{21}].
Hence Eq.~\rqn{25} becomes finally
 \be
\bar{R}_f=\sum_{i}\Pi_{i,w}\bar{R}_{f,i}.
 \e{28}
This equation shows that the pointer expectation value is a classical average of the pointer values corresponding to the eigenvalues of $A$. Hence, $\bar{R}_f$ corresponds to a usual value of $A$ equal to
 \be
A_w=\sum_{i}\Pi_{i,w}a_i.
 \e{A1}
This sum is the spectral representation of the weak value (Eq.~(2.62) in Ref.~\cite{kof12}).

In a similar fashion, we obtain Eqs.~\rqn{25}, \rqn{28}, and \rqn{A1} also when $A$ commutes with $E$.
The difference is only in intermediate formulas.
In particular, now Eq.~\rqn{A5} should be replaced by
 \be
E=\sum_{i,k}e_{ik}|ik\rangle\langle ik|,
 \e{A3}
and Eq.~\rqn{27} should be replaced by
 \be
P_{ik}=\frac{\langle ik|\rho|ik\rangle e_{ik}}{\sum_{i,k}\langle ik|\rho|ik\rangle e_{ik}}.
 \e{A4}

\section{Effects of the system dynamics on PPS measurements of arbitrary strength}
\label{C}

Here we show that the unitary evolution of the quantum system in PPS measurements of arbitrary strength can be taken into account just by redefining the relevant quantities.
We also specify the most general form of the necessary substitutions.

For arbitrary-strength PPS measurements, the expectation value of the pointer variable $R$ of the meter is given by \cite{kof12}
 \be
\bar{R}_s=\frac{{\rm Tr}\,[(E\otimes\hat{R})U_c(A)(\rho\otimes\rho_{\rm M})U_c^\dagger(A)]}{{\rm Tr}\,[(E\otimes I_{\rm M})U_c(A)(\rho\otimes\rho_{\rm M})U_c^\dagger(A)]},
 \e{C1}
where $U_c(A)$ is the correlating operator defined in Appendix \ref{A} and $I_{\rm M}$ is the unity operator for the meter.
Equation \rqn{C1} is obtained for the case of the zero Hamiltonians of the system and the meter.
Here we are interested in the effects of the system dynamics (for the effects of the meter dynamics see Ref.~\cite{kof12} and references therein).

It is easy to see that in the case of impulsive measurements [where the application of the correlating transformation $U_c(A)$ is performed during a very short interval around the moment $t$], the system dynamics yields the substitutions in Eq.~\rqn{C1} given by Eq.~\rqn{57} and
 \be
U_c(A)\rightarrow U(t_1,t)U_c(A),
 \e{C2}
where $U(t_1,t)$ describes the system dynamics [see Eq.~\rqn{58}].

The extra factor $U(t_1,t)$ in Eq.~\rqn{C2} can be included in the definitions of the quantities entering Eq.~\rqn{C1}.
Namely, Eq.~\rqn{C1} with the substitutions \rqn{57} and \rqn{C2} can be recast in the same form as Eq.~\rqn{C1} under the replacements,
\bes{C3'}
 \bea
&&A\rightarrow U_1AU_1^\dagger,\label{C3}\\
&&\rho\rightarrow U_1\rho(t)U_1^\dagger,\label{C4}\\
&&E\rightarrow U_1E(t_1,t)U_1^\dagger.
 \ea{C5}\ese
Here $E(t_1,t)$ is defined in Eq.~\rqn{33}, and $U_1$ is an arbitrary unitary operator.

Equations \rqn{C3'} are the most general transformations of the parameters of the problem, describing the system dynamics for arbitrary-strength PPS measurements.
They are defined with the accuracy to an arbitrary unitary transformation.
Usually in the theory of PPS measurements, the special case of Eqs.~\rqn{C3'} with $U_1=I$ is considered [see Eqs.~\rqn{57'}].
Here we emphasize the fact that this special case is not unique, and other representations are equally allowed by quantum mechanics.
This freedom of choice of the representation may be utilized to simplify and clarify the consideration of PPS measurements.

Note in passing that Eqs.~\rqn{C3'} imply that even in the absence of the system dynamics, PPS measurements are invariant under simultaneous unitary transformations of the relevant parameters,
\bes{C6'}
 \bea
&&A\rightarrow U_1AU_1^\dagger,\label{C6}\\
&&\rho\rightarrow U_1\rho U_1^\dagger,\label{C7}\\
&&E\rightarrow U_1EU_1^\dagger.
 \ea{C8}\ese
This invariance is a special case of the fundamental fact that the results of quantum mechanics are invariant under simultaneous change of all observables and states due to the same unitary transformation.

\bibliography{weakstate}

\begin{thebibliography}{62}%
\makeatletter
\providecommand \@ifxundefined [1]{%
 \@ifx{#1\undefined}
}%
\providecommand \@ifnum [1]{%
 \ifnum #1\expandafter \@firstoftwo
 \else \expandafter \@secondoftwo
 \fi
}%
\providecommand \@ifx [1]{%
 \ifx #1\expandafter \@firstoftwo
 \else \expandafter \@secondoftwo
 \fi
}%
\providecommand \natexlab [1]{#1}%
\providecommand \enquote  [1]{``#1''}%
\providecommand \bibnamefont  [1]{#1}%
\providecommand \bibfnamefont [1]{#1}%
\providecommand \citenamefont [1]{#1}%
\providecommand \href@noop [0]{\@secondoftwo}%
\providecommand \href [0]{\begingroup \@sanitize@url \@href}%
\providecommand \@href[1]{\@@startlink{#1}\@@href}%
\providecommand \@@href[1]{\endgroup#1\@@endlink}%
\providecommand \@sanitize@url [0]{\catcode `\\12\catcode `\$12\catcode
  `\&12\catcode `\#12\catcode `\^12\catcode `\_12\catcode `\%12\relax}%
\providecommand \@@startlink[1]{}%
\providecommand \@@endlink[0]{}%
\providecommand \url  [0]{\begingroup\@sanitize@url \@url }%
\providecommand \@url [1]{\endgroup\@href {#1}{\urlprefix }}%
\providecommand \urlprefix  [0]{URL }%
\providecommand \Eprint [0]{\href }%
\providecommand \doibase [0]{http://dx.doi.org/}%
\providecommand \selectlanguage [0]{\@gobble}%
\providecommand \bibinfo  [0]{\@secondoftwo}%
\providecommand \bibfield  [0]{\@secondoftwo}%
\providecommand \translation [1]{[#1]}%
\providecommand \BibitemOpen [0]{}%
\providecommand \bibitemStop [0]{}%
\providecommand \bibitemNoStop [0]{.\EOS\space}%
\providecommand \EOS [0]{\spacefactor3000\relax}%
\providecommand \BibitemShut  [1]{\csname bibitem#1\endcsname}%
\let\auto@bib@innerbib\@empty
\bibitem [{\citenamefont {Aharonov}\ \emph {et~al.}(1964)\citenamefont
  {Aharonov}, \citenamefont {Bergmann},\ and\ \citenamefont
  {Lebowitz}}]{aha64}%
  \BibitemOpen
  \bibfield  {author} {\bibinfo {author} {\bibfnamefont {Y.}~\bibnamefont
  {Aharonov}}, \bibinfo {author} {\bibfnamefont {P.~G.}\ \bibnamefont
  {Bergmann}}, \ and\ \bibinfo {author} {\bibfnamefont {J.~L.}\ \bibnamefont
  {Lebowitz}},\ }\href@noop {} {\bibfield  {journal} {\bibinfo  {journal}
  {Phys. Rev.}\ }\textbf {\bibinfo {volume} {134}},\ \bibinfo {pages} {B1410}
  (\bibinfo {year} {1964})}\BibitemShut {NoStop}%
\bibitem [{\citenamefont {Tsang}(2009)}]{tsa09}%
  \BibitemOpen
  \bibfield  {author} {\bibinfo {author} {\bibfnamefont {M.}~\bibnamefont
  {Tsang}},\ }\href@noop {} {\bibfield  {journal} {\bibinfo  {journal} {Phys.
  Rev. Lett.}\ }\textbf {\bibinfo {volume} {102}},\ \bibinfo {pages} {250403}
  (\bibinfo {year} {2009})}\BibitemShut {NoStop}%
\bibitem [{\citenamefont {Aharonov}\ \emph {et~al.}(1988)\citenamefont
  {Aharonov}, \citenamefont {Albert},\ and\ \citenamefont {Vaidman}}]{aha88}%
  \BibitemOpen
  \bibfield  {author} {\bibinfo {author} {\bibfnamefont {Y.}~\bibnamefont
  {Aharonov}}, \bibinfo {author} {\bibfnamefont {D.~Z.}\ \bibnamefont
  {Albert}}, \ and\ \bibinfo {author} {\bibfnamefont {L.}~\bibnamefont
  {Vaidman}},\ }\href@noop {} {\bibfield  {journal} {\bibinfo  {journal} {Phys.
  Rev. Lett.}\ }\textbf {\bibinfo {volume} {60}},\ \bibinfo {pages} {1351}
  (\bibinfo {year} {1988})}\BibitemShut {NoStop}%
\bibitem [{\citenamefont {Aharonov}\ and\ \citenamefont
  {Vaidman}(2002)}]{aha02}%
  \BibitemOpen
  \bibfield  {author} {\bibinfo {author} {\bibfnamefont {V.}~\bibnamefont
  {Aharonov}}\ and\ \bibinfo {author} {\bibfnamefont {L.}~\bibnamefont
  {Vaidman}},\ }in\ \href@noop {} {\emph {\bibinfo {booktitle} {Time in Quantum
  Mechanics}}},\ \bibinfo {editor} {edited by\ \bibinfo {editor} {\bibfnamefont
  {J.}~\bibnamefont {Muga}}, \bibinfo {editor} {\bibfnamefont {R.~S.}\
  \bibnamefont {Mayato}}, \ and\ \bibinfo {editor} {\bibfnamefont {I.~L.}\
  \bibnamefont {Egusquiza}}}\ (\bibinfo  {publisher} {Springer},\ \bibinfo
  {year} {2002})\ pp.\ \bibinfo {pages} {399--447}\BibitemShut {NoStop}%
\bibitem [{\citenamefont {Aharonov}\ and\ \citenamefont
  {Rohrlich}(2005)}]{aha05}%
  \BibitemOpen
  \bibfield  {author} {\bibinfo {author} {\bibfnamefont {Y.}~\bibnamefont
  {Aharonov}}\ and\ \bibinfo {author} {\bibfnamefont {D.}~\bibnamefont
  {Rohrlich}},\ }\href@noop {} {\emph {\bibinfo {title} {Quantum Paradoxes}}}\
  (\bibinfo  {publisher} {Wiley-VCH},\ \bibinfo {address} {Weinheim},\ \bibinfo
  {year} {2005})\BibitemShut {NoStop}%
\bibitem [{\citenamefont {Aharonov}\ \emph {et~al.}(2010)\citenamefont
  {Aharonov}, \citenamefont {Popescu},\ and\ \citenamefont
  {Tollaksen}}]{aha10}%
  \BibitemOpen
  \bibfield  {author} {\bibinfo {author} {\bibfnamefont {Y.}~\bibnamefont
  {Aharonov}}, \bibinfo {author} {\bibfnamefont {S.}~\bibnamefont {Popescu}}, \
  and\ \bibinfo {author} {\bibfnamefont {J.}~\bibnamefont {Tollaksen}},\
  }\href@noop {} {\bibfield  {journal} {\bibinfo  {journal} {Phys. Today}\
  }\textbf {\bibinfo {volume} {63}},\ \bibinfo {pages} {27} (\bibinfo {year}
  {2010})}\BibitemShut {NoStop}%
\bibitem [{\citenamefont {Kofman}\ \emph {et~al.}(2012)\citenamefont {Kofman},
  \citenamefont {Ashhab},\ and\ \citenamefont {Nori}}]{kof12}%
  \BibitemOpen
  \bibfield  {author} {\bibinfo {author} {\bibfnamefont {A.~G.}\ \bibnamefont
  {Kofman}}, \bibinfo {author} {\bibfnamefont {S.}~\bibnamefont {Ashhab}}, \
  and\ \bibinfo {author} {\bibfnamefont {F.}~\bibnamefont {Nori}},\ }\href@noop
  {} {\bibfield  {journal} {\bibinfo  {journal} {Phys. Rep.}\ }\textbf
  {\bibinfo {volume} {520}},\ \bibinfo {pages} {43} (\bibinfo {year}
  {2012})}\BibitemShut {NoStop}%
\bibitem [{\citenamefont {Shikano}(2012)}]{shi12}%
  \BibitemOpen
  \bibfield  {author} {\bibinfo {author} {\bibfnamefont {Y.}~\bibnamefont
  {Shikano}},\ }in\ \href@noop {} {\emph {\bibinfo {booktitle} {Measurements in
  Quantum Mechanics}}},\ \bibinfo {editor} {edited by\ \bibinfo {editor}
  {\bibfnamefont {M.~R.}\ \bibnamefont {Pahlavani}}}\ (\bibinfo  {publisher}
  {InTech},\ \bibinfo {year} {2012})\ pp.\ \bibinfo {pages}
  {75--100}\BibitemShut {NoStop}%
\bibitem [{\citenamefont {Dressel}\ \emph {et~al.}(2013)\citenamefont
  {Dressel}, \citenamefont {Malik}, \citenamefont {Miatto}, \citenamefont
  {Jordan},\ and\ \citenamefont {Boyd}}]{dre}%
  \BibitemOpen
  \bibfield  {author} {\bibinfo {author} {\bibfnamefont {J.}~\bibnamefont
  {Dressel}}, \bibinfo {author} {\bibfnamefont {M.}~\bibnamefont {Malik}},
  \bibinfo {author} {\bibfnamefont {F.~M.}\ \bibnamefont {Miatto}}, \bibinfo
  {author} {\bibfnamefont {A.~N.}\ \bibnamefont {Jordan}}, \ and\ \bibinfo
  {author} {\bibfnamefont {R.~W.}\ \bibnamefont {Boyd}},\ }\href@noop {}
  {\bibfield  {journal} {\bibinfo  {journal} {arXiv:1305.7154}\ } (\bibinfo
  {year} {2013})}\BibitemShut {NoStop}%
\bibitem [{\citenamefont {Aharonov}\ and\ \citenamefont
  {Vaidman}(1991)}]{aha91}%
  \BibitemOpen
  \bibfield  {author} {\bibinfo {author} {\bibfnamefont {Y.}~\bibnamefont
  {Aharonov}}\ and\ \bibinfo {author} {\bibfnamefont {L.}~\bibnamefont
  {Vaidman}},\ }\href@noop {} {\bibfield  {journal} {\bibinfo  {journal} {J.
  Phys. A}\ }\textbf {\bibinfo {volume} {24}},\ \bibinfo {pages} {2315}
  (\bibinfo {year} {1991})}\BibitemShut {NoStop}%
\bibitem [{\citenamefont {Resch}\ \emph {et~al.}(2004)\citenamefont {Resch},
  \citenamefont {Lundeen},\ and\ \citenamefont {Steinberg}}]{res04}%
  \BibitemOpen
  \bibfield  {author} {\bibinfo {author} {\bibfnamefont {K.~J.}\ \bibnamefont
  {Resch}}, \bibinfo {author} {\bibfnamefont {J.~S.}\ \bibnamefont {Lundeen}},
  \ and\ \bibinfo {author} {\bibfnamefont {A.~M.}\ \bibnamefont {Steinberg}},\
  }\href@noop {} {\bibfield  {journal} {\bibinfo  {journal} {Phys. Lett. A}\
  }\textbf {\bibinfo {volume} {324}},\ \bibinfo {pages} {125} (\bibinfo {year}
  {2004})}\BibitemShut {NoStop}%
\bibitem [{\citenamefont {Aharonov}\ \emph {et~al.}(2002)\citenamefont
  {Aharonov}, \citenamefont {Botero}, \citenamefont {Popescu}, \citenamefont
  {Reznik},\ and\ \citenamefont {Tollaksen}}]{aha02a}%
  \BibitemOpen
  \bibfield  {author} {\bibinfo {author} {\bibfnamefont {Y.}~\bibnamefont
  {Aharonov}}, \bibinfo {author} {\bibfnamefont {A.}~\bibnamefont {Botero}},
  \bibinfo {author} {\bibfnamefont {S.}~\bibnamefont {Popescu}}, \bibinfo
  {author} {\bibfnamefont {B.}~\bibnamefont {Reznik}}, \ and\ \bibinfo {author}
  {\bibfnamefont {J.}~\bibnamefont {Tollaksen}},\ }\href@noop {} {\bibfield
  {journal} {\bibinfo  {journal} {Phys. Lett. A}\ }\textbf {\bibinfo {volume}
  {301}},\ \bibinfo {pages} {130} (\bibinfo {year} {2002})}\BibitemShut
  {NoStop}%
\bibitem [{\citenamefont {Lundeen}\ and\ \citenamefont
  {Steinberg}(2009)}]{lun09}%
  \BibitemOpen
  \bibfield  {author} {\bibinfo {author} {\bibfnamefont {J.~S.}\ \bibnamefont
  {Lundeen}}\ and\ \bibinfo {author} {\bibfnamefont {A.~M.}\ \bibnamefont
  {Steinberg}},\ }\href@noop {} {\bibfield  {journal} {\bibinfo  {journal}
  {Phys. Rev. Lett.}\ }\textbf {\bibinfo {volume} {102}},\ \bibinfo {pages}
  {020404} (\bibinfo {year} {2009})}\BibitemShut {NoStop}%
\bibitem [{\citenamefont {Yokota}\ \emph {et~al.}(2009)\citenamefont {Yokota},
  \citenamefont {Yamamoto}, \citenamefont {Koashi},\ and\ \citenamefont
  {Imoto}}]{yok09}%
  \BibitemOpen
  \bibfield  {author} {\bibinfo {author} {\bibfnamefont {K.}~\bibnamefont
  {Yokota}}, \bibinfo {author} {\bibfnamefont {T.}~\bibnamefont {Yamamoto}},
  \bibinfo {author} {\bibfnamefont {M.}~\bibnamefont {Koashi}}, \ and\ \bibinfo
  {author} {\bibfnamefont {N.}~\bibnamefont {Imoto}},\ }\href@noop {}
  {\bibfield  {journal} {\bibinfo  {journal} {New J. Phys.}\ }\textbf {\bibinfo
  {volume} {11}},\ \bibinfo {pages} {033011} (\bibinfo {year}
  {2009})}\BibitemShut {NoStop}%
\bibitem [{\citenamefont {Goggin}\ \emph {et~al.}(2011)\citenamefont {Goggin},
  \citenamefont {Almeida}, \citenamefont {Barbieri}, \citenamefont {Lanyon},
  \citenamefont {O'Bryen}, \citenamefont {White},\ and\ \citenamefont
  {Pryde}}]{gog11}%
  \BibitemOpen
  \bibfield  {author} {\bibinfo {author} {\bibfnamefont {M.~E.}\ \bibnamefont
  {Goggin}}, \bibinfo {author} {\bibfnamefont {M.~P.}\ \bibnamefont {Almeida}},
  \bibinfo {author} {\bibfnamefont {M.}~\bibnamefont {Barbieri}}, \bibinfo
  {author} {\bibfnamefont {B.~P.}\ \bibnamefont {Lanyon}}, \bibinfo {author}
  {\bibfnamefont {J.~L.}\ \bibnamefont {O'Bryen}}, \bibinfo {author}
  {\bibfnamefont {A.~G.}\ \bibnamefont {White}}, \ and\ \bibinfo {author}
  {\bibfnamefont {G.~J.}\ \bibnamefont {Pryde}},\ }\href@noop {} {\bibfield
  {journal} {\bibinfo  {journal} {Proc. Nat. Acad. Sci.}\ }\textbf {\bibinfo
  {volume} {108}},\ \bibinfo {pages} {1256} (\bibinfo {year}
  {2011})}\BibitemShut {NoStop}%
\bibitem [{\citenamefont {Lundeen}\ \emph {et~al.}(2011)\citenamefont
  {Lundeen}, \citenamefont {Sutherland}, \citenamefont {Patel}, \citenamefont
  {Stewart},\ and\ \citenamefont {Bamber}}]{lun11}%
  \BibitemOpen
  \bibfield  {author} {\bibinfo {author} {\bibfnamefont {J.~S.}\ \bibnamefont
  {Lundeen}}, \bibinfo {author} {\bibfnamefont {B.}~\bibnamefont {Sutherland}},
  \bibinfo {author} {\bibfnamefont {A.}~\bibnamefont {Patel}}, \bibinfo
  {author} {\bibfnamefont {C.}~\bibnamefont {Stewart}}, \ and\ \bibinfo
  {author} {\bibfnamefont {C.}~\bibnamefont {Bamber}},\ }\href@noop {}
  {\bibfield  {journal} {\bibinfo  {journal} {Nature}\ }\textbf {\bibinfo
  {volume} {474}},\ \bibinfo {pages} {188} (\bibinfo {year}
  {2011})}\BibitemShut {NoStop}%
\bibitem [{\citenamefont {Salavail}\ \emph {et~al.}(2013)\citenamefont
  {Salavail}, \citenamefont {Agnew}, \citenamefont {Johnson}, \citenamefont
  {Bolduc}, \citenamefont {Leach},\ and\ \citenamefont {Boyd}}]{sal13}%
  \BibitemOpen
  \bibfield  {author} {\bibinfo {author} {\bibfnamefont {J.~Z.}\ \bibnamefont
  {Salavail}}, \bibinfo {author} {\bibfnamefont {M.}~\bibnamefont {Agnew}},
  \bibinfo {author} {\bibfnamefont {A.~S.}\ \bibnamefont {Johnson}}, \bibinfo
  {author} {\bibfnamefont {E.}~\bibnamefont {Bolduc}}, \bibinfo {author}
  {\bibfnamefont {J.}~\bibnamefont {Leach}}, \ and\ \bibinfo {author}
  {\bibfnamefont {R.~W.}\ \bibnamefont {Boyd}},\ }\href@noop {} {\bibfield
  {journal} {\bibinfo  {journal} {Nature Photon.}\ }\textbf {\bibinfo {volume}
  {7}},\ \bibinfo {pages} {316} (\bibinfo {year} {2013})}\BibitemShut {NoStop}%
\bibitem [{\citenamefont {Kocsis}\ \emph {et~al.}(2011)\citenamefont {Kocsis},
  \citenamefont {Braverman}, \citenamefont {Ravets}, \citenamefont {Stevens},
  \citenamefont {Mirin}, \citenamefont {Shalm},\ and\ \citenamefont
  {Steinberg}}]{koc11}%
  \BibitemOpen
  \bibfield  {author} {\bibinfo {author} {\bibfnamefont {S.}~\bibnamefont
  {Kocsis}}, \bibinfo {author} {\bibfnamefont {B.}~\bibnamefont {Braverman}},
  \bibinfo {author} {\bibfnamefont {S.}~\bibnamefont {Ravets}}, \bibinfo
  {author} {\bibfnamefont {M.~J.}\ \bibnamefont {Stevens}}, \bibinfo {author}
  {\bibfnamefont {R.~P.}\ \bibnamefont {Mirin}}, \bibinfo {author}
  {\bibfnamefont {L.~K.}\ \bibnamefont {Shalm}}, \ and\ \bibinfo {author}
  {\bibfnamefont {A.~M.}\ \bibnamefont {Steinberg}},\ }\href@noop {} {\bibfield
   {journal} {\bibinfo  {journal} {Science}\ }\textbf {\bibinfo {volume}
  {332}},\ \bibinfo {pages} {1170} (\bibinfo {year} {2011})}\BibitemShut
  {NoStop}%
\bibitem [{\citenamefont {Bliokh}\ \emph {et~al.}(2013)\citenamefont {Bliokh},
  \citenamefont {Bekshaev}, \citenamefont {Kofman},\ and\ \citenamefont
  {Nori}}]{bli13}%
  \BibitemOpen
  \bibfield  {author} {\bibinfo {author} {\bibfnamefont {K.~Y.}\ \bibnamefont
  {Bliokh}}, \bibinfo {author} {\bibfnamefont {A.~Y.}\ \bibnamefont
  {Bekshaev}}, \bibinfo {author} {\bibfnamefont {A.~G.}\ \bibnamefont
  {Kofman}}, \ and\ \bibinfo {author} {\bibfnamefont {F.}~\bibnamefont
  {Nori}},\ }\href@noop {} {\bibfield  {journal} {\bibinfo  {journal} {New J.
  Phys.}\ }\textbf {\bibinfo {volume} {15}},\ \bibinfo {pages} {073022}
  (\bibinfo {year} {2013})}\BibitemShut {NoStop}%
\bibitem [{\citenamefont {Hosten}\ and\ \citenamefont {Kwiat}(2008)}]{hos08}%
  \BibitemOpen
  \bibfield  {author} {\bibinfo {author} {\bibfnamefont {O.}~\bibnamefont
  {Hosten}}\ and\ \bibinfo {author} {\bibfnamefont {P.}~\bibnamefont {Kwiat}},\
  }\href@noop {} {\bibfield  {journal} {\bibinfo  {journal} {Science}\ }\textbf
  {\bibinfo {volume} {319}},\ \bibinfo {pages} {787} (\bibinfo {year}
  {2008})}\BibitemShut {NoStop}%
\bibitem [{\citenamefont {Dixon}\ \emph {et~al.}(2009)\citenamefont {Dixon},
  \citenamefont {Starling}, \citenamefont {Jordan},\ and\ \citenamefont
  {Howell}}]{dix09}%
  \BibitemOpen
  \bibfield  {author} {\bibinfo {author} {\bibfnamefont {P.~B.}\ \bibnamefont
  {Dixon}}, \bibinfo {author} {\bibfnamefont {D.~J.}\ \bibnamefont {Starling}},
  \bibinfo {author} {\bibfnamefont {A.~N.}\ \bibnamefont {Jordan}}, \ and\
  \bibinfo {author} {\bibfnamefont {J.~C.}\ \bibnamefont {Howell}},\
  }\href@noop {} {\bibfield  {journal} {\bibinfo  {journal} {Phys. Rev. Lett.}\
  }\textbf {\bibinfo {volume} {102}},\ \bibinfo {pages} {173601} (\bibinfo
  {year} {2009})}\BibitemShut {NoStop}%
\bibitem [{\citenamefont {Zhou}\ \emph {et~al.}(2013)\citenamefont {Zhou},
  \citenamefont {Turek}, \citenamefont {Sun},\ and\ \citenamefont
  {Nori}}]{zho}%
  \BibitemOpen
  \bibfield  {author} {\bibinfo {author} {\bibfnamefont {L.}~\bibnamefont
  {Zhou}}, \bibinfo {author} {\bibfnamefont {Y.}~\bibnamefont {Turek}},
  \bibinfo {author} {\bibfnamefont {C.~P.}\ \bibnamefont {Sun}}, \ and\
  \bibinfo {author} {\bibfnamefont {F.}~\bibnamefont {Nori}},\ }\href@noop {}
  {\bibfield  {journal} {\bibinfo  {journal} {arXiv:1302.0455}\ } (\bibinfo
  {year} {2013})}\BibitemShut {NoStop}%
\bibitem [{\citenamefont {Gorodetski}\ \emph {et~al.}(2012)\citenamefont
  {Gorodetski}, \citenamefont {Bliokh}, \citenamefont {Stein}, \citenamefont
  {Genet}, \citenamefont {Shitrit}, \citenamefont {Kleiner}, \citenamefont
  {Hasman},\ and\ \citenamefont {Ebbesen}}]{gor12}%
  \BibitemOpen
  \bibfield  {author} {\bibinfo {author} {\bibfnamefont {Y.}~\bibnamefont
  {Gorodetski}}, \bibinfo {author} {\bibfnamefont {K.~Y.}\ \bibnamefont
  {Bliokh}}, \bibinfo {author} {\bibfnamefont {B.}~\bibnamefont {Stein}},
  \bibinfo {author} {\bibfnamefont {C.}~\bibnamefont {Genet}}, \bibinfo
  {author} {\bibfnamefont {N.}~\bibnamefont {Shitrit}}, \bibinfo {author}
  {\bibfnamefont {V.}~\bibnamefont {Kleiner}}, \bibinfo {author} {\bibfnamefont
  {E.}~\bibnamefont {Hasman}}, \ and\ \bibinfo {author} {\bibfnamefont {T.~W.}\
  \bibnamefont {Ebbesen}},\ }\href@noop {} {\bibfield  {journal} {\bibinfo
  {journal} {Phys. Rev. Lett.}\ }\textbf {\bibinfo {volume} {109}},\ \bibinfo
  {pages} {013901} (\bibinfo {year} {2012})}\BibitemShut {NoStop}%
\bibitem [{\citenamefont {Leggett}(1989)}]{leg89}%
  \BibitemOpen
  \bibfield  {author} {\bibinfo {author} {\bibfnamefont {A.~J.}\ \bibnamefont
  {Leggett}},\ }\href@noop {} {\bibfield  {journal} {\bibinfo  {journal} {Phys.
  Rev. Lett.}\ }\textbf {\bibinfo {volume} {62}},\ \bibinfo {pages} {2325}
  (\bibinfo {year} {1989})}\BibitemShut {NoStop}%
\bibitem [{\citenamefont {Peres}(1989)}]{per89}%
  \BibitemOpen
  \bibfield  {author} {\bibinfo {author} {\bibfnamefont {A.}~\bibnamefont
  {Peres}},\ }\href@noop {} {\bibfield  {journal} {\bibinfo  {journal} {Phys.
  Rev. Lett.}\ }\textbf {\bibinfo {volume} {62}},\ \bibinfo {pages} {2326}
  (\bibinfo {year} {1989})}\BibitemShut {NoStop}%
\bibitem [{\citenamefont {Svensson}(2013)}]{sve}%
  \BibitemOpen
  \bibfield  {author} {\bibinfo {author} {\bibfnamefont {B.~E.~Y.}\
  \bibnamefont {Svensson}},\ }\href@noop {} {\bibfield  {journal} {\bibinfo
  {journal} {arXiv:1301.4328}\ } (\bibinfo {year} {2013})}\BibitemShut
  {NoStop}%
\bibitem [{\citenamefont {Berry}\ and\ \citenamefont {Shukla}(2010)}]{ber10}%
  \BibitemOpen
  \bibfield  {author} {\bibinfo {author} {\bibfnamefont {M.~V.}\ \bibnamefont
  {Berry}}\ and\ \bibinfo {author} {\bibfnamefont {P.}~\bibnamefont {Shukla}},\
  }\href@noop {} {\bibfield  {journal} {\bibinfo  {journal} {J. Phys. A}\
  }\textbf {\bibinfo {volume} {43}},\ \bibinfo {pages} {354024} (\bibinfo
  {year} {2010})}\BibitemShut {NoStop}%
\bibitem [{\citenamefont {Berry}\ \emph {et~al.}(2011)\citenamefont {Berry},
  \citenamefont {Dennis}, \citenamefont {McRoberts},\ and\ \citenamefont
  {Shukla}}]{ber11}%
  \BibitemOpen
  \bibfield  {author} {\bibinfo {author} {\bibfnamefont {M.~V.}\ \bibnamefont
  {Berry}}, \bibinfo {author} {\bibfnamefont {M.~R.}\ \bibnamefont {Dennis}},
  \bibinfo {author} {\bibfnamefont {B.}~\bibnamefont {McRoberts}}, \ and\
  \bibinfo {author} {\bibfnamefont {P.}~\bibnamefont {Shukla}},\ }\href@noop {}
  {\bibfield  {journal} {\bibinfo  {journal} {J. Phys. A}\ }\textbf {\bibinfo
  {volume} {44}},\ \bibinfo {pages} {205301} (\bibinfo {year}
  {2011})}\BibitemShut {NoStop}%
\bibitem [{\citenamefont {Aharonov}\ \emph {et~al.}()\citenamefont {Aharonov},
  \citenamefont {Popescu},\ and\ \citenamefont {Skrzypczyk}}]{aha}%
  \BibitemOpen
  \bibfield  {author} {\bibinfo {author} {\bibfnamefont {Y.}~\bibnamefont
  {Aharonov}}, \bibinfo {author} {\bibfnamefont {S.}~\bibnamefont {Popescu}}, \
  and\ \bibinfo {author} {\bibfnamefont {P.}~\bibnamefont {Skrzypczyk}},\
  }\href@noop {} {\bibinfo  {journal} {arxiv:1202.0631}\ }\BibitemShut
  {NoStop}%
\bibitem [{\citenamefont {Shikano}\ and\ \citenamefont {Hosoya}(2010)}]{shi10}%
  \BibitemOpen
\bibfield  {journal} {  }\bibfield  {author} {\bibinfo {author} {\bibfnamefont
  {Y.}~\bibnamefont {Shikano}}\ and\ \bibinfo {author} {\bibfnamefont
  {A.}~\bibnamefont {Hosoya}},\ }\href@noop {} {\bibfield  {journal} {\bibinfo
  {journal} {J. Phys. A}\ }\textbf {\bibinfo {volume} {43}},\ \bibinfo {pages}
  {025304} (\bibinfo {year} {2010})}\BibitemShut {NoStop}%
\bibitem [{\citenamefont {Hosoya}\ and\ \citenamefont {Koga}(2011)}]{hos11}%
  \BibitemOpen
  \bibfield  {author} {\bibinfo {author} {\bibfnamefont {A.}~\bibnamefont
  {Hosoya}}\ and\ \bibinfo {author} {\bibfnamefont {M.}~\bibnamefont {Koga}},\
  }\href@noop {} {\bibfield  {journal} {\bibinfo  {journal} {J. Phys. A}\
  }\textbf {\bibinfo {volume} {44}},\ \bibinfo {pages} {415303} (\bibinfo
  {year} {2011})}\BibitemShut {NoStop}%
\bibitem [{\citenamefont {Hofmann}(2010)}]{hof10}%
  \BibitemOpen
  \bibfield  {author} {\bibinfo {author} {\bibfnamefont {H.~F.}\ \bibnamefont
  {Hofmann}},\ }\href@noop {} {\bibfield  {journal} {\bibinfo  {journal} {Phys.
  Rev. A}\ }\textbf {\bibinfo {volume} {81}},\ \bibinfo {pages} {012103}
  (\bibinfo {year} {2010})}\BibitemShut {NoStop}%
\bibitem [{\citenamefont {Kofman}\ \emph {et~al.}(2013)\citenamefont {Kofman},
  \citenamefont {{\"{O}zdemir}},\ and\ \citenamefont {Nori}}]{kof13}%
  \BibitemOpen
  \bibfield  {author} {\bibinfo {author} {\bibfnamefont {A.~G.}\ \bibnamefont
  {Kofman}}, \bibinfo {author} {\bibfnamefont {{\c{S}}.~K.}\ \bibnamefont
  {{\"{O}zdemir}}}, \ and\ \bibinfo {author} {\bibfnamefont {F.}~\bibnamefont
  {Nori}},\ }\href@noop {} {\bibfield  {journal} {\bibinfo  {journal}
  {arXiv:1303.6031}\ } (\bibinfo {year} {2013})}\BibitemShut {NoStop}%
\bibitem [{\citenamefont {Hiroishi}\ and\ \citenamefont
  {Hofmann}(2013)}]{hir13}%
  \BibitemOpen
  \bibfield  {author} {\bibinfo {author} {\bibfnamefont {M.}~\bibnamefont
  {Hiroishi}}\ and\ \bibinfo {author} {\bibfnamefont {H.~F.}\ \bibnamefont
  {Hofmann}},\ }\href@noop {} {\bibfield  {journal} {\bibinfo  {journal} {J.
  Phys. A}\ }\textbf {\bibinfo {volume} {46}},\ \bibinfo {pages} {245302}
  (\bibinfo {year} {2013})}\BibitemShut {NoStop}%
\bibitem [{\citenamefont {Barnett}\ \emph {et~al.}(2000)\citenamefont
  {Barnett}, \citenamefont {Pegg},\ and\ \citenamefont {Jeffers}}]{bar00}%
  \BibitemOpen
  \bibfield  {author} {\bibinfo {author} {\bibfnamefont {S.~M.}\ \bibnamefont
  {Barnett}}, \bibinfo {author} {\bibfnamefont {D.~T.}\ \bibnamefont {Pegg}}, \
  and\ \bibinfo {author} {\bibfnamefont {J.}~\bibnamefont {Jeffers}},\
  }\href@noop {} {\bibfield  {journal} {\bibinfo  {journal} {J. Mod. Opt.}\
  }\textbf {\bibinfo {volume} {47}},\ \bibinfo {pages} {1779} (\bibinfo {year}
  {2000})}\BibitemShut {NoStop}%
\bibitem [{\citenamefont {von Neumann}(1955)}]{neu55}%
  \BibitemOpen
  \bibfield  {author} {\bibinfo {author} {\bibfnamefont {J.}~\bibnamefont {von
  Neumann}},\ }\href@noop {} {\emph {\bibinfo {title} {Mathematical Foundations
  of Quantum Mechanics}}}\ (\bibinfo  {publisher} {Princeton University},\
  \bibinfo {address} {Princeton, NJ},\ \bibinfo {year} {1955})\BibitemShut
  {NoStop}%
\bibitem [{\citenamefont {L\"{u}ders}(1951)}]{lud51}%
  \BibitemOpen
  \bibfield  {author} {\bibinfo {author} {\bibfnamefont {G.}~\bibnamefont
  {L\"{u}ders}},\ }\href@noop {} {\bibfield  {journal} {\bibinfo  {journal}
  {Ann. Phys. (Leipzig)}\ }\textbf {\bibinfo {volume} {8}},\ \bibinfo {pages}
  {322} (\bibinfo {year} {1951})}\BibitemShut {NoStop}%
\bibitem [{\citenamefont {Nielsen}\ and\ \citenamefont {Chuang}(2000)}]{nie00}%
  \BibitemOpen
  \bibfield  {author} {\bibinfo {author} {\bibfnamefont {M.~A.}\ \bibnamefont
  {Nielsen}}\ and\ \bibinfo {author} {\bibfnamefont {I.~L.}\ \bibnamefont
  {Chuang}},\ }\href@noop {} {\emph {\bibinfo {title} {Quantum Computation and
  Quantum Information}}}\ (\bibinfo  {publisher} {Cambridge University Press},\
  \bibinfo {year} {2000})\BibitemShut {NoStop}%
\bibitem [{\citenamefont {Dressel}\ and\ \citenamefont {Jordan}(2012)}]{dre12}%
  \BibitemOpen
  \bibfield  {author} {\bibinfo {author} {\bibfnamefont {J.}~\bibnamefont
  {Dressel}}\ and\ \bibinfo {author} {\bibfnamefont {A.~N.}\ \bibnamefont
  {Jordan}},\ }\href@noop {} {\bibfield  {journal} {\bibinfo  {journal} {Phys.
  Rev. A}\ }\textbf {\bibinfo {volume} {85}},\ \bibinfo {pages} {022123}
  (\bibinfo {year} {2012})}\BibitemShut {NoStop}%
\bibitem [{\citenamefont {Wiseman}(2002)}]{wis02}%
  \BibitemOpen
  \bibfield  {author} {\bibinfo {author} {\bibfnamefont {H.~M.}\ \bibnamefont
  {Wiseman}},\ }\href@noop {} {\bibfield  {journal} {\bibinfo  {journal} {Phys.
  Rev. A}\ }\textbf {\bibinfo {volume} {65}},\ \bibinfo {pages} {032111}
  (\bibinfo {year} {2002})}\BibitemShut {NoStop}%
\bibitem [{\citenamefont {Johansen}\ and\ \citenamefont {Luis}(2004)}]{joh04}%
  \BibitemOpen
  \bibfield  {author} {\bibinfo {author} {\bibfnamefont {L.~M.}\ \bibnamefont
  {Johansen}}\ and\ \bibinfo {author} {\bibfnamefont {A.}~\bibnamefont
  {Luis}},\ }\href@noop {} {\bibfield  {journal} {\bibinfo  {journal} {Phys.
  Rev. A}\ }\textbf {\bibinfo {volume} {70}},\ \bibinfo {pages} {052115}
  (\bibinfo {year} {2004})}\BibitemShut {NoStop}%
\bibitem [{\citenamefont {Bub}\ and\ \citenamefont {Brown}(1986)}]{bub86}%
  \BibitemOpen
  \bibfield  {author} {\bibinfo {author} {\bibfnamefont {J.}~\bibnamefont
  {Bub}}\ and\ \bibinfo {author} {\bibfnamefont {H.}~\bibnamefont {Brown}},\
  }\href@noop {} {\bibfield  {journal} {\bibinfo  {journal} {Phys. Rev. Lett.}\
  }\textbf {\bibinfo {volume} {56}},\ \bibinfo {pages} {2337} (\bibinfo {year}
  {1986})}\BibitemShut {NoStop}%
\bibitem [{Note1()}]{Note1}%
  \BibitemOpen
  \bibinfo {note} {In the case of pure pre- and post-selected states,
  unnormalized connection states were considered in Ref.~\cite
  {shi10}}\BibitemShut {NoStop}%
\bibitem [{Note2()}]{Note2}%
  \BibitemOpen
  \bibinfo {note} {Here we assume that $A$ is Hermitian. Though sometimes weak
  values of non-Hermitian operators are discussed \cite {wis02}, here for
  simplicity we do not consider such weak values}\BibitemShut {NoStop}%
\bibitem [{\citenamefont {Wiseman}\ and\ \citenamefont
  {Milburn}(2010)}]{wis10}%
  \BibitemOpen
  \bibfield  {author} {\bibinfo {author} {\bibfnamefont {H.~M.}\ \bibnamefont
  {Wiseman}}\ and\ \bibinfo {author} {\bibfnamefont {G.~J.}\ \bibnamefont
  {Milburn}},\ }\href@noop {} {\emph {\bibinfo {title} {Quantum Measurements
  and Control}}}\ (\bibinfo  {publisher} {Cambridge University Press},\
  \bibinfo {year} {2010})\BibitemShut {NoStop}%
\bibitem [{Note3()}]{Note3}%
  \BibitemOpen
  \bibinfo {note} {The same conclusion was obtained from different
  considerations in Ref.~\cite {ste95}}\BibitemShut {NoStop}%
\bibitem [{Note4()}]{Note4}%
  \BibitemOpen
  \bibinfo {note} {The Hermitian operator $w'$ was considered previously in
  Ref.~\cite {hof10}, where it was called ``the transient density
  matrix''}\BibitemShut {NoStop}%
\bibitem [{\citenamefont {Franklin}(2000)}]{fra00}%
  \BibitemOpen
  \bibfield  {author} {\bibinfo {author} {\bibfnamefont {J.~N.}\ \bibnamefont
  {Franklin}},\ }\href@noop {} {\emph {\bibinfo {title} {Matrix Theory}}}\
  (\bibinfo  {publisher} {Dover},\ \bibinfo {address} {Mineola, NY},\ \bibinfo
  {year} {2000})\BibitemShut {NoStop}%
\bibitem [{\citenamefont {Steinberg}(1995)}]{ste95}%
  \BibitemOpen
  \bibfield  {author} {\bibinfo {author} {\bibfnamefont {A.~M.}\ \bibnamefont
  {Steinberg}},\ }\href@noop {} {\bibfield  {journal} {\bibinfo  {journal}
  {Phys. Rev. Lett.}\ }\textbf {\bibinfo {volume} {74}},\ \bibinfo {pages}
  {2405} (\bibinfo {year} {1995})}\BibitemShut {NoStop}%
\bibitem [{\citenamefont {Aharonov}\ and\ \citenamefont
  {Botero}(2005)}]{aha05a}%
  \BibitemOpen
  \bibfield  {author} {\bibinfo {author} {\bibfnamefont {Y.}~\bibnamefont
  {Aharonov}}\ and\ \bibinfo {author} {\bibfnamefont {A.}~\bibnamefont
  {Botero}},\ }\href@noop {} {\bibfield  {journal} {\bibinfo  {journal} {Phys.
  Rev. A}\ }\textbf {\bibinfo {volume} {72}},\ \bibinfo {pages} {052111}
  (\bibinfo {year} {2005})}\BibitemShut {NoStop}%
\bibitem [{Note5()}]{Note5}%
  \BibitemOpen
  \bibinfo {note} {This statement with $\rho _{\protect \rm eff}=w'$ was proved
  in Ref.~\cite {kof12}, Sec.~14.2. The case $\rho _{\protect \rm eff}=w$ is
  proved similarly, with the difference that now in Eq.~(14.6) in Ref.~\cite
  {kof12} the relation (D.7), and not (D.8), should be used}\BibitemShut
  {NoStop}%
\bibitem [{\citenamefont {Aharonov}\ and\ \citenamefont
  {Vaidman}(1990)}]{aha90}%
  \BibitemOpen
  \bibfield  {author} {\bibinfo {author} {\bibfnamefont {Y.}~\bibnamefont
  {Aharonov}}\ and\ \bibinfo {author} {\bibfnamefont {L.}~\bibnamefont
  {Vaidman}},\ }\href@noop {} {\bibfield  {journal} {\bibinfo  {journal} {Phys.
  Rev. A}\ }\textbf {\bibinfo {volume} {41}},\ \bibinfo {pages} {11} (\bibinfo
  {year} {1990})}\BibitemShut {NoStop}%
\bibitem [{\citenamefont {Hofmann}\ and\ \citenamefont
  {Takeuchi}(2003)}]{hof03}%
  \BibitemOpen
  \bibfield  {author} {\bibinfo {author} {\bibfnamefont {H.~F.}\ \bibnamefont
  {Hofmann}}\ and\ \bibinfo {author} {\bibfnamefont {S.}~\bibnamefont
  {Takeuchi}},\ }\href@noop {} {\bibfield  {journal} {\bibinfo  {journal}
  {Phys. Rev. A}\ }\textbf {\bibinfo {volume} {68}},\ \bibinfo {pages} {032103}
  (\bibinfo {year} {2003})}\BibitemShut {NoStop}%
\bibitem [{\citenamefont {Amri}\ \emph {et~al.}(2011)\citenamefont {Amri},
  \citenamefont {Laurat},\ and\ \citenamefont {Fabre}}]{amr11}%
  \BibitemOpen
  \bibfield  {author} {\bibinfo {author} {\bibfnamefont {T.}~\bibnamefont
  {Amri}}, \bibinfo {author} {\bibfnamefont {J.}~\bibnamefont {Laurat}}, \ and\
  \bibinfo {author} {\bibfnamefont {C.}~\bibnamefont {Fabre}},\ }\href@noop {}
  {\bibfield  {journal} {\bibinfo  {journal} {Phys. Rev. Lett.}\ }\textbf
  {\bibinfo {volume} {106}},\ \bibinfo {pages} {020502} (\bibinfo {year}
  {2011})}\BibitemShut {NoStop}%
\bibitem [{\citenamefont {Lundeen}\ \emph {et~al.}(2009)\citenamefont
  {Lundeen}, \citenamefont {Feito}, \citenamefont {Coldenstrodt-Ronge},
  \citenamefont {Pregnell}, \citenamefont {Silberhorn}, \citenamefont {Ralph},
  \citenamefont {Eisert}, \citenamefont {Plenio},\ and\ \citenamefont
  {Walmsley}}]{lun09a}%
  \BibitemOpen
  \bibfield  {author} {\bibinfo {author} {\bibfnamefont {J.~S.}\ \bibnamefont
  {Lundeen}}, \bibinfo {author} {\bibfnamefont {A.}~\bibnamefont {Feito}},
  \bibinfo {author} {\bibfnamefont {H.}~\bibnamefont {Coldenstrodt-Ronge}},
  \bibinfo {author} {\bibfnamefont {K.~L.}\ \bibnamefont {Pregnell}}, \bibinfo
  {author} {\bibfnamefont {C.}~\bibnamefont {Silberhorn}}, \bibinfo {author}
  {\bibfnamefont {T.~C.}\ \bibnamefont {Ralph}}, \bibinfo {author}
  {\bibfnamefont {J.}~\bibnamefont {Eisert}}, \bibinfo {author} {\bibfnamefont
  {M.~B.}\ \bibnamefont {Plenio}}, \ and\ \bibinfo {author} {\bibfnamefont
  {I.~A.}\ \bibnamefont {Walmsley}},\ }\href@noop {} {\bibfield  {journal}
  {\bibinfo  {journal} {Nature Phys.}\ }\textbf {\bibinfo {volume} {5}},\
  \bibinfo {pages} {27} (\bibinfo {year} {2009})}\BibitemShut {NoStop}%
\bibitem [{\citenamefont {Balian}\ and\ \citenamefont
  {V\'{e}n\'{e}roni}(1993)}]{bal93}%
  \BibitemOpen
  \bibfield  {author} {\bibinfo {author} {\bibfnamefont {R.}~\bibnamefont
  {Balian}}\ and\ \bibinfo {author} {\bibfnamefont {M.}~\bibnamefont
  {V\'{e}n\'{e}roni}},\ }\href@noop {} {\bibfield  {journal} {\bibinfo
  {journal} {Nucl. Phys. B}\ }\textbf {\bibinfo {volume} {408}},\ \bibinfo
  {pages} {445} (\bibinfo {year} {1993})}\BibitemShut {NoStop}%
\bibitem [{\citenamefont {Pegg}(2008)}]{peg08}%
  \BibitemOpen
  \bibfield  {author} {\bibinfo {author} {\bibfnamefont {D.~T.}\ \bibnamefont
  {Pegg}},\ }\href@noop {} {\bibfield  {journal} {\bibinfo  {journal} {Found.
  Phys.}\ }\textbf {\bibinfo {volume} {38}},\ \bibinfo {pages} {648} (\bibinfo
  {year} {2008})}\BibitemShut {NoStop}%
\bibitem [{\citenamefont {Barnett}\ and\ \citenamefont {Pegg}(1999)}]{bar99}%
  \BibitemOpen
  \bibfield  {author} {\bibinfo {author} {\bibfnamefont {S.~M.}\ \bibnamefont
  {Barnett}}\ and\ \bibinfo {author} {\bibfnamefont {D.~T.}\ \bibnamefont
  {Pegg}},\ }\href@noop {} {\bibfield  {journal} {\bibinfo  {journal} {Phys.
  Rev. A}\ }\textbf {\bibinfo {volume} {60}},\ \bibinfo {pages} {4965}
  (\bibinfo {year} {1999})}\BibitemShut {NoStop}%
\bibitem [{Note6()}]{Note6}%
  \BibitemOpen
  \bibinfo {note} {In particular, in Refs.~\cite {bar99,bar00} the first two
  expressions in Eq.~(\ref {53}) are called the ``retrodictive formalism'' in
  quantum mechanics and the third expression is called the ``predictive
  formalism'', whereas in Ref.~\cite {peg08} a relation to causality in quantum
  mechanics was considered}\BibitemShut {NoStop}%
\bibitem [{\citenamefont {Aharonov}\ \emph {et~al.}(2009)\citenamefont
  {Aharonov}, \citenamefont {Popescu}, \citenamefont {Tollaksen},\ and\
  \citenamefont {Vaidman}}]{aha09}%
  \BibitemOpen
  \bibfield  {author} {\bibinfo {author} {\bibfnamefont {Y.}~\bibnamefont
  {Aharonov}}, \bibinfo {author} {\bibfnamefont {S.}~\bibnamefont {Popescu}},
  \bibinfo {author} {\bibfnamefont {J.}~\bibnamefont {Tollaksen}}, \ and\
  \bibinfo {author} {\bibfnamefont {L.}~\bibnamefont {Vaidman}},\ }\href@noop
  {} {\bibfield  {journal} {\bibinfo  {journal} {Phys. Rev. A}\ }\textbf
  {\bibinfo {volume} {79}},\ \bibinfo {pages} {052110} (\bibinfo {year}
  {2009})}\BibitemShut {NoStop}%
\bibitem [{\citenamefont {Korotkov}(1999)}]{kor99}%
  \BibitemOpen
  \bibfield  {author} {\bibinfo {author} {\bibfnamefont {A.~N.}\ \bibnamefont
  {Korotkov}},\ }\href@noop {} {\bibfield  {journal} {\bibinfo  {journal}
  {Phys. Rev. B}\ }\textbf {\bibinfo {volume} {60}},\ \bibinfo {pages} {5737}
  (\bibinfo {year} {1999})}\BibitemShut {NoStop}%
\bibitem [{\citenamefont {Gardiner}\ and\ \citenamefont
  {Zoller}(2004)}]{gar04}%
  \BibitemOpen
  \bibfield  {author} {\bibinfo {author} {\bibfnamefont {C.~W.}\ \bibnamefont
  {Gardiner}}\ and\ \bibinfo {author} {\bibfnamefont {P.}~\bibnamefont
  {Zoller}},\ }\href@noop {} {\emph {\bibinfo {title} {Quantum Noise}}}\
  (\bibinfo  {publisher} {Springer},\ \bibinfo {address} {Berlin},\ \bibinfo
  {year} {2004})\BibitemShut {NoStop}%
\end{thebibliography}%



\end{document}